\documentclass[
 reprint,
superscriptaddress,
amsmath,amssymb,
prl,
]{revtex4-2}

\usepackage{aas_macros}
\usepackage{graphicx}% Include figure files
\usepackage{dcolumn}% Align table columns on decimal point
\usepackage{bm}% bold math
\usepackage{hyperref}% add hypertext capabilities
\usepackage{color}
\usepackage[mathlines]{lineno}% Enable numbering of text and display math
\usepackage{tabularx}
\usepackage{subfigure}
\usepackage[normalem]{ulem}
\usepackage[export]{adjustbox}
\usepackage{url}
\usepackage{lineno}
\usepackage[flushleft]{threeparttable}
\usepackage{makecell}
\usepackage{multirow}
\usepackage{rotating}

\hypersetup{
    pdfnewwindow=true,    % links in new window
    colorlinks=true,      % false: boxed links; true: colored links
    linkcolor=blue,       % color of internal links
    citecolor=blue,       % color of links to bibliography
    filecolor=blue,       % color of file links
    urlcolor=blue         % color of external links
}

\def\orcid#1{\href{https://orcid.org/#1}{\includegraphics[keepaspectratio,width=1.1em]{Fig./orcid.png}}}
\usepackage{natbib} % for super citation

\begin{document}

\title{Constraints on Ultra Heavy Dark Matter Properties from Dwarf Spheroidal Galaxies with LHAASO Observations}% Force line breaks with \\

\author{Zhen Cao}
\affiliation{Key Laboratory of Particle Astrophysics \& Experimental Physics Division \& Computing Center, Institute of High Energy Physics, Chinese Academy of Sciences, 100049 Beijing, China}
\affiliation{University of Chinese Academy of Sciences, 100049 Beijing, China}
\affiliation{Tianfu Cosmic Ray Research Center, 610000 Chengdu, Sichuan,  China}
 
\author{F. Aharonian}
\affiliation{Dublin Institute for Advanced Studies, 31 Fitzwilliam Place, 2 Dublin, Ireland }
\affiliation{Max-Planck-Institut for Nuclear Physics, P.O. Box 103980, 69029  Heidelberg, Germany}
 
\author{Q. An}
\affiliation{State Key Laboratory of Particle Detection and Electronics, China}
\affiliation{University of Science and Technology of China, 230026 Hefei, Anhui, China}
 
\author{Axikegu}
\affiliation{School of Physical Science and Technology \&  School of Information Science and Technology, Southwest Jiaotong University, 610031 Chengdu, Sichuan, China}
 
\author{Y.X. Bai}
\affiliation{Key Laboratory of Particle Astrophysics \& Experimental Physics Division \& Computing Center, Institute of High Energy Physics, Chinese Academy of Sciences, 100049 Beijing, China}
\affiliation{Tianfu Cosmic Ray Research Center, 610000 Chengdu, Sichuan,  China}
 
\author{Y.W. Bao}
\affiliation{School of Astronomy and Space Science, Nanjing University, 210023 Nanjing, Jiangsu, China}
 
\author{D. Bastieri}
\affiliation{Center for Astrophysics, Guangzhou University, 510006 Guangzhou, Guangdong, China}
 
\author{X.J. Bi }
\affiliation{Key Laboratory of Particle Astrophysics \& Experimental Physics Division \& Computing Center, Institute of High Energy Physics, Chinese Academy of Sciences, 100049 Beijing, China}
\affiliation{University of Chinese Academy of Sciences, 100049 Beijing, China}
\affiliation{Tianfu Cosmic Ray Research Center, 610000 Chengdu, Sichuan,  China}
 
\author{Y.J. Bi}
\affiliation{Key Laboratory of Particle Astrophysics \& Experimental Physics Division \& Computing Center, Institute of High Energy Physics, Chinese Academy of Sciences, 100049 Beijing, China}
\affiliation{Tianfu Cosmic Ray Research Center, 610000 Chengdu, Sichuan,  China}
 
\author{J.T. Cai}
\affiliation{Center for Astrophysics, Guangzhou University, 510006 Guangzhou, Guangdong, China}
 
% ...

\collaboration{The LHAASO Collaboration}
\altaffiliation[Corresponding Author:]{\\ jiakang@mail.sdu.edu.cn (K. Jia) \\ lij@pmo.ac.cn (J. Li) \\ bixj@ihep.ac.cn (X.J. Bi) \\ gaolinqing@hotmail.com (L.Q. Gao) \\ xyhuang@pmo.ac.cn (X.Y. Huang) \\ liwl@mail.sdu.edu.cn (W.L. Li) \\ zhucg@email.sdu.edu.cn (C.G. Zhu)}

\date{\today}% It is always \today, today,
             %  but any date may be explicitly specified

\begin{abstract}
In this work we try to search for signals generated by ultra-heavy dark matter at the Large High Altitude Air Shower Observatory (LHAASO) data. We look for possible gamma-ray by dark matter annihilation or decay from 16 dwarf spheroidal galaxies in the field of view of LHAASO. Dwarf spheroidal galaxies are among the most promising targets for indirect detection of dark matter which have low fluxes of astrophysical $\gamma$-ray background while large amount of dark matter. By analyzing more than 700 days observational data at LHAASO, no significant dark matter signal from 1 TeV to 1 EeV is detected. Accordingly we derive the most stringent constraints on the ultra-heavy dark matter annihilation cross-section up to EeV. The constraints on the lifetime of dark matter in decay mode are also derived. 
\end{abstract}

\maketitle
%TC:endignore
%\tableofcontents
\textit{Introduction}---
Various kinds of astronomical evidence suggest the existence of massive dark matter (DM) in the universe~\cite{Bertone_2005}, which comprises approximately 85$\%$ of all matter~\cite{Planck:2018vyg}. However, DM cannot be explained by the Standard Model of particle physics~\cite{JUNGMAN_1996,Bergstr_m_2012}. Therefore, one of the most important tasks in fundamental physics is to detect and reveal the nature of DM particles.
{Most searches primarily focus on weakly interacting massive particles (WIMPs) or
ultra-light DM. However, no conclusive DM signal has been observed up to now~\cite{DAMPE:2021hsz,An:2022hhb,Foster:2022fxn,LZ:2022lsv,XENON:2023cxc}. 
On the other hand, ultra-heavy dark matter (UHDM; 10 TeV $\lesssim M_{\chi}\lesssim m_{pl}\approx10^{16}$ TeV) represents a potential alternative DM candidate that could be generated through various mechanisms, including freeze-out, freeze-in, out-of-equilibrium decay, phase transitions, gravitational particle production, and primordial black holes (see the review
in Ref.~\cite{Carney_2023} and references therein). Some models for UHDM, like composite dark matter~\cite{Harigaya_2016,Geller2018}, have been proposed to evade the unitarity limit, and very-high-energy (VHE) gamma-ray may be produced not only via the decay of UHDM, but also via its self-annihilation~\cite{Tak_2022}.}

Among different astronomical systems, dwarf spheroidal galaxies (dSphs) are considered one of the most promising targets for detecting DM signals due to their relatively short distances, high mass-to-light ratios \cite{Strigari_2013,Conrad_2017}, and locations far away from complicated emission regions like the Galactic disk. These properties have instigated extensive research on them utilizing various astronomical facilities~\cite{Fermi-LAT:2011vow, Fermi_2015,HAWC_2018,HAWC_2020,H.E.S.S_2020,MAGIC_2022,VERITAS_2023,Guo:2022rqq}. Importantly, given the relative proximity of these systems, the angular dimensions of their signal regions, particularly in scenarios involving decay, may be comparable to or even surpass the point spread function (PSF) of detection instruments. Thus, viewing these sources as extended rather than point-like sources may play a crucial role in the indirect detection of DM~\cite{Fermi-LAT:2012fij, DiMauro:2022hue}.

The LHAASO is located in Sichuan Province, China, at an altitude of 4410 meters. It is a multi-purpose and comprehensive extensive air shower array, designed for the study of cosmic rays and gamma-ray across wide energy ranges, from 10 TeV to 100 PeV for cosmic rays and from sub-TeV to beyond 1 PeV for gamma-ray \cite{Ma_2022}. {LHAASO is composed of three sub-arrays: the  KiloMeter Squared Array (KM2A), the Water Cherenkov Detector Array (WCDA), and the Wide Field-of-view air Cherenkov Telescopes Array (WFCTA).} Since its operation, several important results have been achieved in cosmic-ray and gamma-ray research \cite{LHAASO_nature,PhysRevLett.128.051102,KM2A_GCHalo_2022,PhysRevLett.126.241103,LHAASO_GRB2023}. The remarkable gamma-ray sensitivity of LHAASO for energies exceeding 100 TeV~\cite{ZHEN2019457} presents an opportunity for the exploration of UHDM. {WCDA and KM2A also have good PSFs for VHE gamma-ray~\cite{Aharonian_2021_a,Aharonian_2021}, enabling them to potentially discern the spatial extension of dSphs.}

In this Letter, we search for VHE $\gamma$-ray signal from dSphs with data recorded by WCDA and KM2A of LHAASO, and report the stringent constraints on the UHDM up to EeV.

\textit{Gamma-ray Flux from Dark Matter}---The expected differential gamma-ray flux from DM annihilation can be written as

\begin{equation}
\label{FluxAnni}
\begin{aligned}
\frac{{\rm d}F_{\rm anni}}{{\rm d}E{\rm d}\Omega {\rm d}t}(E,\Omega) = \frac{\left\langle \sigma_Av \right\rangle}{8\pi M^2_{\chi}} \frac{{\rm d}N_{\gamma}}{{\rm d}E} e^{-\tau_{\gamma \gamma}(E)} \times \frac{{\rm d}J}{{\rm d}\Omega}.
\end{aligned}
\end{equation}
Similarly, for DM decay, it can be defined as

\begin{equation}
\label{FluxDecay}
\begin{aligned}
\frac{{\rm d}F_{\rm decay}}{{\rm d}E{\rm d}\Omega {\rm d}t}(E,\Omega) = \frac{1}{4\pi \tau_{\chi} M_{\chi}} \frac{{\rm d}N_{\gamma}}{{\rm d}E} e^{-\tau_{\gamma \gamma}(E)} \times \frac{{\rm d}D}{{\rm d}\Omega},
\end{aligned}
\end{equation}
where $\left\langle \sigma_Av \right\rangle$ is the velocity-weighted DM annihilation cross-section, $\tau_{\chi}$ is the DM decay lifetime, and $M_{\chi}$ is the mass of the DM particle. ${\rm d}N_\gamma/{\rm d}E$ is the gamma-ray energy spectrum resulting from DM annihilation (decay), as calculated using HDMSpectra~\cite{HDMS_2021}. The term $\tau_{\gamma \gamma}(E)$ represents the total attenuation depth resulting from the pair production process ($\gamma \gamma \rightarrow e^+e^-$), taking into account background photons from starlight (SL), infrared radiation (IR), and cosmic microwave background (CMB), as described in Ref.~\cite{ISRF_2015}. The last term is the differential J- (D-) factor, which characterizes the strength of the DM signal. In Eq.~\ref{FluxAnni} and Eq.~\ref{FluxDecay}, ${\rm d}J/{\rm d}\Omega = \int \rho^{2}_{DM}(r) {\rm d}l$, and ${\rm d}D/{\rm d}\Omega = \int \rho_{DM}(r) {\rm d}l$, where $\rho_{DM}(r)$ refers to the DM density at distance $r$ from the center of dSphs, and $l$ represents the distance from a point on the line-of-sight (L.o.S) to the Earth. The J- (D-) factor is defined as the differential J- (D-) factor integrated over the region of interest (ROI). In this work, we take the DM density distribution in dSphs following the Navarro-Frenk-White (NFW) profile~\cite{DMProfile_1997}.

%In Eq.~\ref{FluxAnni} and Eq.~\ref{FluxDecay}, $ {\rm d}J/{\rm d}\Omega$ and $ {\rm d}D/{\rm d}\Omega$ are defined as, 
%
%\begin{equation}
%            \begin{aligned}
%            { \frac{{\rm d}J}{{\rm d}\Omega}=\int\rho^{2}_{DM}(r){\rm d}l}, 
%            \label{con:jfactor}
%            \end{aligned}
%            \end{equation}
%\begin{equation}
%            \begin{aligned}
%            { \frac{{\rm d}D}{{\rm d}\Omega}=\int\rho_{DM}(r){\rm d}l}, 
%            \label{con:dfactor}
%            \end{aligned}
%\end{equation}
%where ${\rho_{DM}(r)}$ refers to the DM density at distance $r$ from the center of dSphs, and $l$ represents the distance from a point on the line-of-sight (L.o.S) to the Earth. \textcolor{blue}{And the J-(D-) factor is defined as the differential J-(D-) factor integrated over the region of interest (ROI).} In this work, we take the DM density distribution in dSphs following the Navarro-Frenk-White (NFW) proﬁle\cite{DMProfile_1997}.

\begin{table}[!htbp]%

    \tiny
    \begin{minipage}{0.5\textwidth}
    
	\caption{The ROI half width and J- (D-) factor for 16 dSphs considered in this analysis.} 
	%\centering 
	%\renewcommand\tabcolsep{10.0pt}
    \centering
    \resizebox{\textwidth}{!}{
    \label{tabJD} 
    \begin{tabular}{@{}ccccc@{}}
    \botrule
			 			Name    &$\rm{log_{10}(J_{\theta}/GeV^{2}cm^{-5})}$   &
			 			$\rm{\theta_{anni}[deg]}$  &
			 			$\rm{log_{10}(D_{\theta}/GeVcm^{-2})}$   &
			 			$\rm{\theta_{decay}[deg]}$\\
			 			
			 			\hline
Draco & $18.96_{-0.15}^{+0.16}$ & 1.0 & $19.38_{-0.32}^{+0.24}$ & 2.3 \\
Ursa Minor & $18.79_{-0.11}^{+0.12}$ & 1.0 & $18.68_{-0.15}^{+0.33}$ & 2.1 \\
Ursa Major I & $18.40_{-0.27}^{+0.28}$ & 0.9 & $18.64_{-0.48}^{+0.50}$ & 1.8 \\
Ursa Major II & $19.70_{-0.43}^{+0.43}$ & 1.0 & $19.41_{-0.57}^{+0.43}$ & 2.0 \\
Bootes 1 & $18.39_{-0.37}^{+0.36}$ & 0.9 & $18.77_{-0.54}^{+0.40}$ & 1.8 \\
Canes Venatici I & $17.43_{-0.15}^{+0.16}$ & 0.8 & $18.19_{-0.39}^{+0.40}$ & 1.3 \\
Coma Berenices & $19.26_{-0.43}^{+0.35}$ & 0.9 & $19.12_{-0.73}^{+0.46}$ & 1.8 \\
Leo I & $17.58_{-0.10}^{+0.10}$ & 0.8 & $18.44_{-0.42}^{+0.33}$ & 1.4 \\
Segue 1 & $19.25_{-0.69}^{+0.60}$ & 0.8 & $18.33_{-0.63}^{+0.69}$ & 0.8 \\
Sextans & $17.80_{-0.10}^{+0.10}$ & 1.0 & $18.49_{-0.21}^{+0.28}$ & 1.8 \\
Canes Venatici II & $17.82_{-0.37}^{+0.38}$ & 0.8 & $18.45_{-0.74}^{+0.50}$ & 1.4 \\
Hercules & $17.60_{-0.69}^{+0.53}$ & 0.8 & $17.79_{-0.61}^{+0.62}$ & 1.0 \\
Leo II & $17.72_{-0.17}^{+0.18}$ & 0.8 & $17.85_{-0.40}^{+0.62}$ & 1.0 \\
Willman I & $19.80_{-0.52}^{+0.50}$ & 0.9 & $19.00_{-0.93}^{+0.71}$ & 1.5 \\
Aquarius 2 & $18.57_{-0.57}^{+0.50}$ & 1.1 & $18.53_{-0.68}^{+0.61}$ & 1.3 \\
Leo T & $17.66_{-0.52}^{+0.55}$ & 0.8 & $17.88_{-0.69}^{+0.65}$ & 1.0 \\
		
			 	\botrule
            \end{tabular}
            }
            \end{minipage}
 \end{table}

With a large field-of-view (FoV) of approximately 2 sr, LHAASO has the ability to observe about 60$\%$ of the sky each day~\cite{ZHEN2019457}. 
The 16 dSphs within the FoV of LHAASO have been selected as our observation targets, and the coordinates of these dSphs are shown in Table~S1 of the Supplemental Material~\cite{Supplement_Materials}. To optimize the size of the ROI for balancing the preference between a larger area containing more signal and a smaller area with less background (and nearby sources) contamination, we utilize $\mathcal{S}/\sqrt{B}$ as a metric, where $\mathcal{S}$ and $B$ are the expected signal and expected background in the ROI, respectively. Considering the expected signal also depends on the details of NFW profile~\cite{Geringer_Sameth_2015,Bonnivard_2015,Pace_2018}, {we use the publicly
available MCMC chains provided by Ref.~\cite{Pace_2018} to determine the optimal ROI for our instrument and compute the corresponding J-(D-) factor distribution in our ROI.} The details are discussed in Sec.~II of the Supplemental Material. The half-width of the chosen ROI, the corresponding median J- (D-) factor, and its uncertainty for each dSph are shown in Table \ref{tabJD}. For the observation of WCDA, it is a conservative case that the ROI selection is consistent with the above value because WCDA has better angular resolution in the low energy range than KM2A. Meanwhile, the J- (D-) factor is consistent throughout the analysis.

\textit{Observation and Data Analysis}---{The present work utilizes LHAASO-WCDA data (E $<$ 20 TeV ) acquired from March 5, 2021 to March 31, 2023. LHAASO-KM2A data (E $>$ 10 TeV) are utilized, including KM2A 1/2 array data from December 27, 2019 to November 30, 2020,  KM2A 3/4 array data from December 1, 2020 to July 19, 2021, and KM2A full array data from July 20, 2021 to February 28, 2022. We apply the detector simulation, event reconstruction and selection algorithms detailed in the performance papers of LHAASO sub-arrays~\cite{Aharonian_2021_a,Aharonian_2021} for the analysis of WCDA and KM2A data. The total effective observation time for each target dSph from WCDA and KM2A are shown in Table~S1 of Supplemental Material~\cite{Supplement_Materials}. }

We divide the KM2A data from 10 TeV to 10$^{3}$ TeV into 10 logarithmically evenly spaced bins according to reconstructed energy. For the WCDA data, events are divided into 6 groups according to the number of triggered PMT units ($N_{hits}$), i.e. [60,100], [100,200], [200,300], [300,500], [500,800], [800,2000].
Based on the reconstructed direction, the selected events from each energy bin in the KM2A dataset and each group of WCDA data are mapped onto a 2D sky map with a pixel size of  0.1$^{\circ} \times$0.1$^{\circ}$ in the equatorial coordinate. We use the ``direct integration" method as described in Ref.~\cite{Fleysher_2004} to estimate the number of  background events per pixel.
{To eliminate the contamination of known gamma-ray sources on background estimation, we mask the Galactic disk region ($|\rm{b}| < 10^{\circ}$), known sources given by TeVCat~\cite{Wakely:2007qpa}, and first LHAASO catalog~\cite{LHAASO:2023rpg} (see Fig.~S1 of Supplemental Material~\cite{Supplement_Materials}).}

The expected numbers of gamma-ray events produced by DM in the ROIs are calculated by folding the gamma-ray flux produced by DM with the WCDA and KM2A detector response function respectively.
More details are discussed in Sec.~I of Supplemental Material~\cite{Supplement_Materials}.

%\textit{Statistics Method}---
To quantify the excess of gamma-ray signals in the ROIs, we use a 3D binned likelihood ratio analysis combining WCDA and KM2A data. This method accounts for both the energy spectrum and the spatial characteristics of the DM signals, which are different from the background in the ROIs. 
In this analysis, we define the 3D likelihood function for the $k$-th dSph as follows: 

\begin{equation}
            \begin{aligned}
             { \mathcal L}_k=\prod \limits_{i,j}{\rm Poisson}(N_{i,j,k}^{obs};N_{i,j,k}^{exp}+N_{i,j,k}^{bkg})\times
            { \mathcal G}(B_k;B_{k}^{obs},\sigma_k)
            ,
            \label{likelihood}
            \end{aligned}
            \end{equation}
where

\begin{equation}
            \begin{split}
             { \mathcal G}(B_k;B_{k}^{obs},\sigma_k)=\frac{1}{{\rm ln}(10)B_{k}^{obs}\sqrt{2\pi}\sigma_k}\\
              \times e^{-[{\rm log_{10}}(B_k)-{\rm log_{10}}(B_{k}^{obs})]^2/2\sigma_k^2}.
            \label{likelihood_G}
            \end{split}
            \end{equation}
The $N_{i,j,k}^{exp}$ is the expected number of gamma-ray from DM annihilation or decay in the $i$-th energy estimator bin and the $j$-th pixel on the 2D sky map of the $k$-th dSph. $N_{i,j,k}^{bkg}$ is the estimated background events from the ``direct integration" method, and $N_{i,j,k}^{obs}$ is the observed number of gamma-ray photons. The term $\mathcal G(B_k;B_{k}^{obs},\sigma_k)$ is included for the statistical uncertainties on the  J- (D-) factor of the $k$-th dSph, following Ref.~\cite{Fermi-LAT:2011vow,Fermi_2015}, where $B$ equals $J$ for the annihilation case and $B$ equals $D$ for the decaying case. The larger uncertainties listed in Table~\ref{tabJD} are taken as $\sigma_k$ considering the asymmetric distribution of J- (D-) factor conservatively.

To quantify how well the DM signal fits the observed data, we define the test statistic of the $k$-th dSph (${TS_k}$) as,

\begin{equation}
            \begin{aligned}
            { TS_k=-2\rm{ln}(\frac{{\mathcal L_k}(S=0)}{{\mathcal L_k}(S_{max})})} ,
            \label{Ts}
            \end{aligned}
            \end{equation}
where ${S}$ represents the DM signal flux, and $S_{\text{max}}$ is the best-fit value of the DM signal flux that maximizes the likelihood. 
To avoid non-physical values, we set  $\left \langle \sigma_Av \right \rangle$ and $\tau_{\chi}$ to be positive during the fitting process. We obtained the statistical significance of the signal over the null hypothesis (no DM model) by $\sqrt{TS_{k}}$. 
Then one-sided 95\% confidence level (C.L.) limits on $ \left \langle \sigma_Av \right \rangle$ or $\tau_{\chi}$ are set by increasing the DM signal normalization from its best-fit value until -2${\rm ln}{\mathcal L}$ increases by a value of 2.71~\cite{Cowan_2011}.
            
The combined likelihood analysis of all dSphs is performed by $\mathcal {L}_{total}=\prod \limits_{k}{\mathcal L}_{k}$, with the aim of improving the overall statistical power and generating stronger constraints {on the DM parameters.}

\begin{figure*}[htbp]
\includegraphics[width=0.45\textwidth]{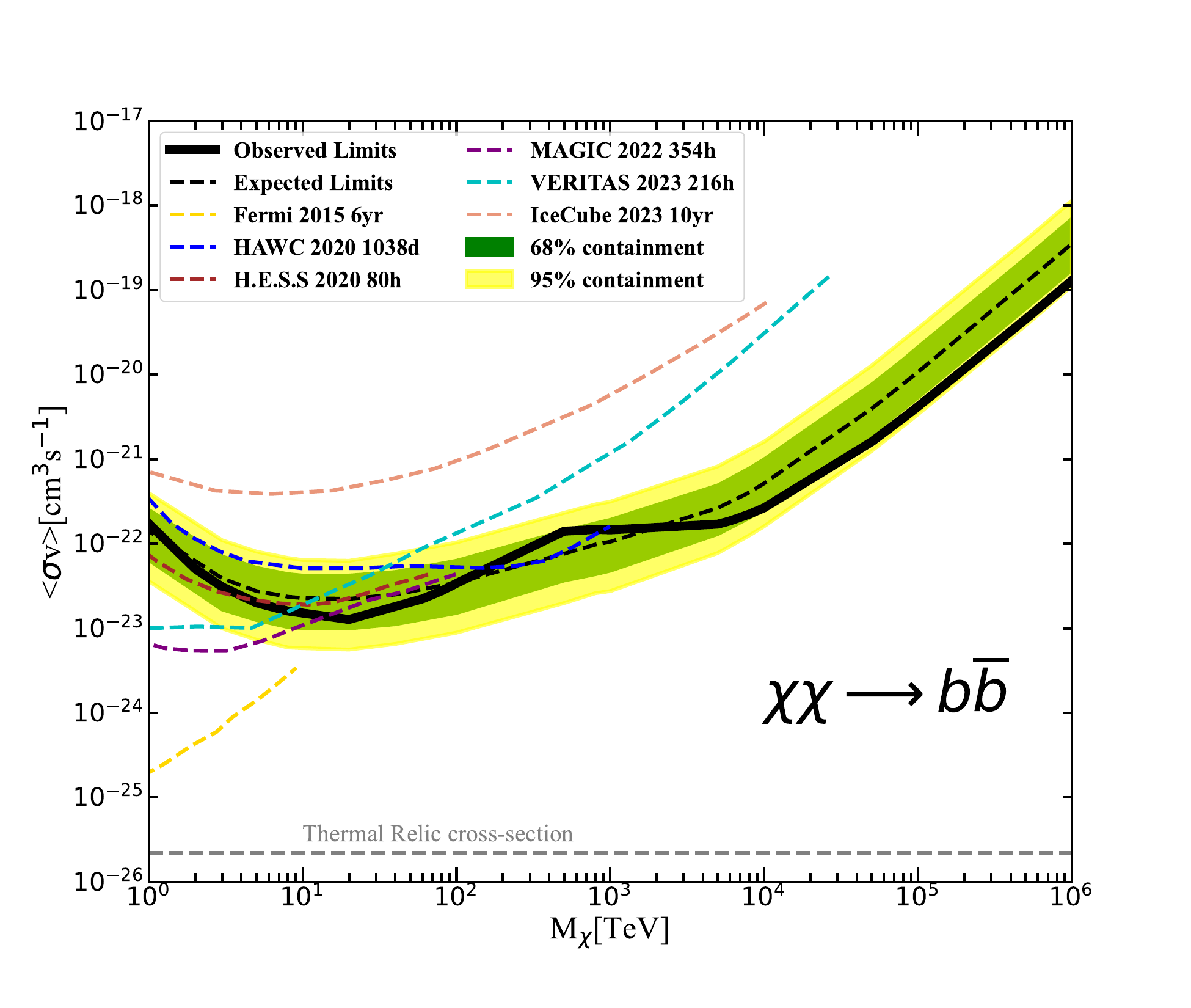}
\includegraphics[width=0.45\textwidth]{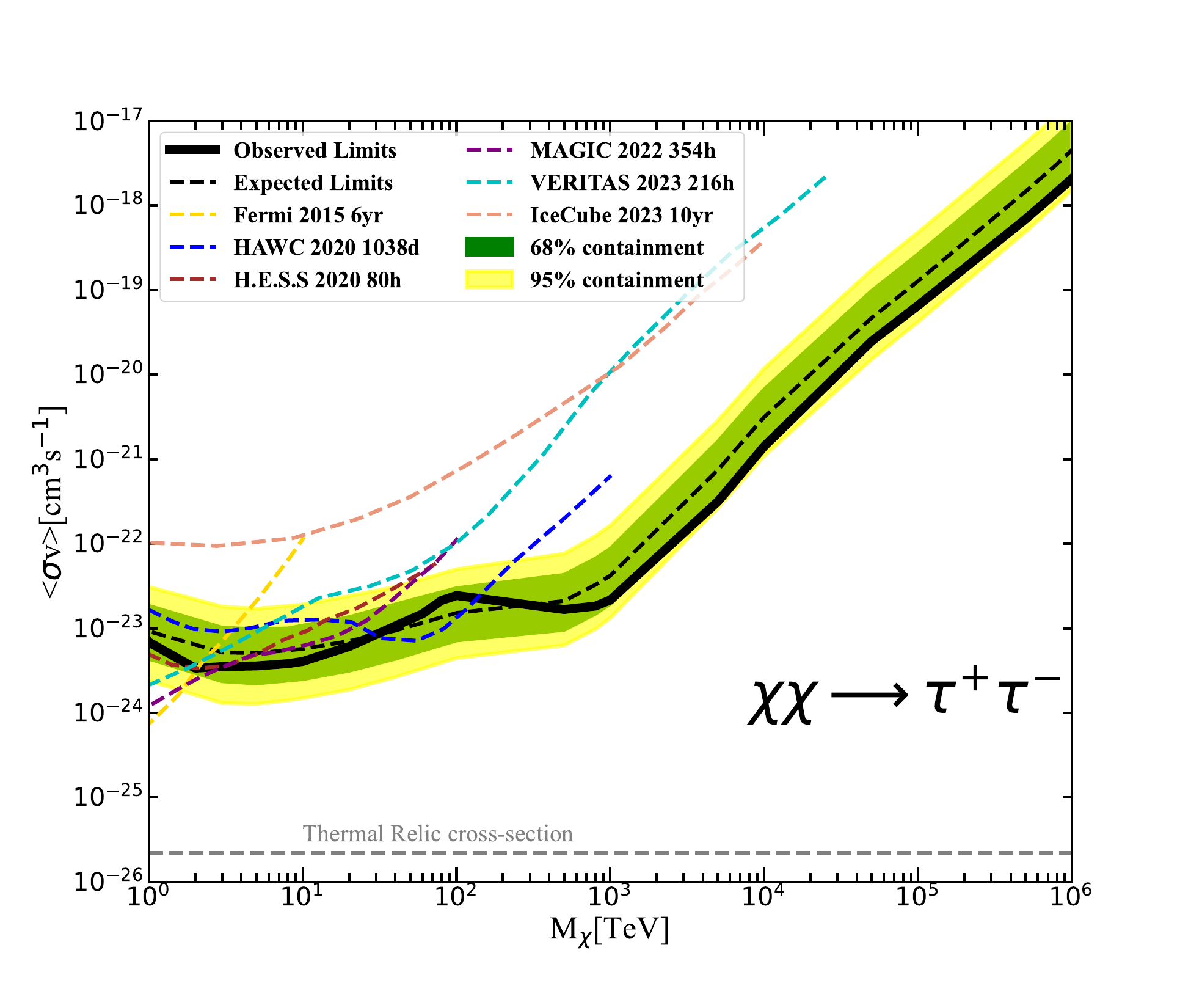}
\caption{The 95$\%$ C.L. upper limits on the DM annihilation cross-section for  $b\overline{b}, \tau^{+}\tau^{-}$ channels and comparing to other experiments (Fermi-LAT~\cite{Fermi_2015}, HAWC~\cite{HAWC_2020} ,H.E.S.S~\cite{H.E.S.S_2020}, MAGIC~\cite{MAGIC_2022}, VERITAS~\cite{VERITAS_2023}, IceCube~\cite{PhysRevD.108.043001}). The solid black line represents the observed combined limit of this work. The dashed black line, green band, and yellow band represent the expected limits and their 1$\sigma$ and 2$\sigma$ uncertainties. The dashed gray line is Thermal Relic cross-section~\cite{PhysRevD.86.023506}, and the other dashed colored lines show the results of other experiments.}
\label{expectlimits_anni}
\end{figure*}

\begin{figure*}[htbp]
\includegraphics[width=0.45\textwidth]{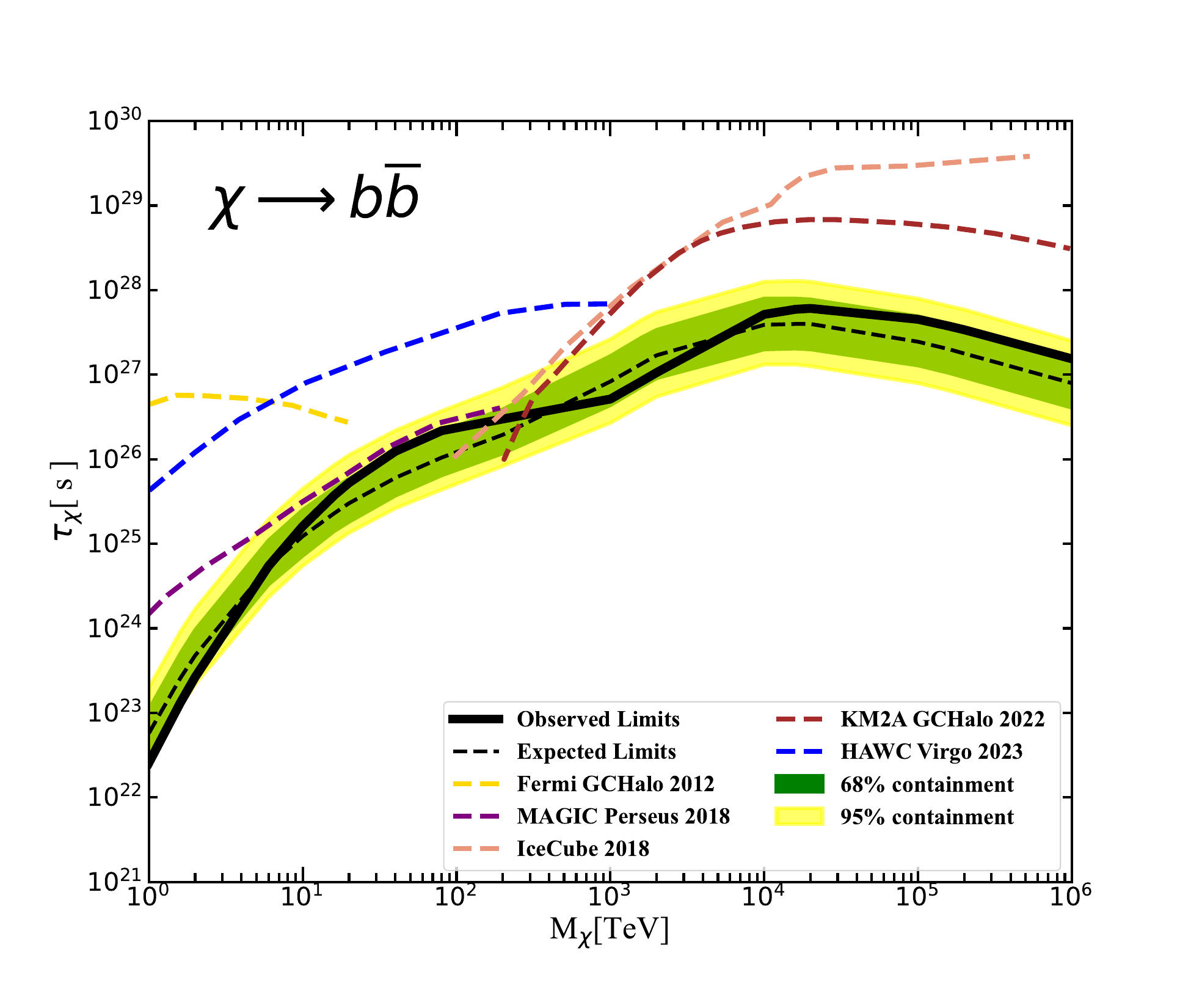}
\includegraphics[width=0.45\textwidth]{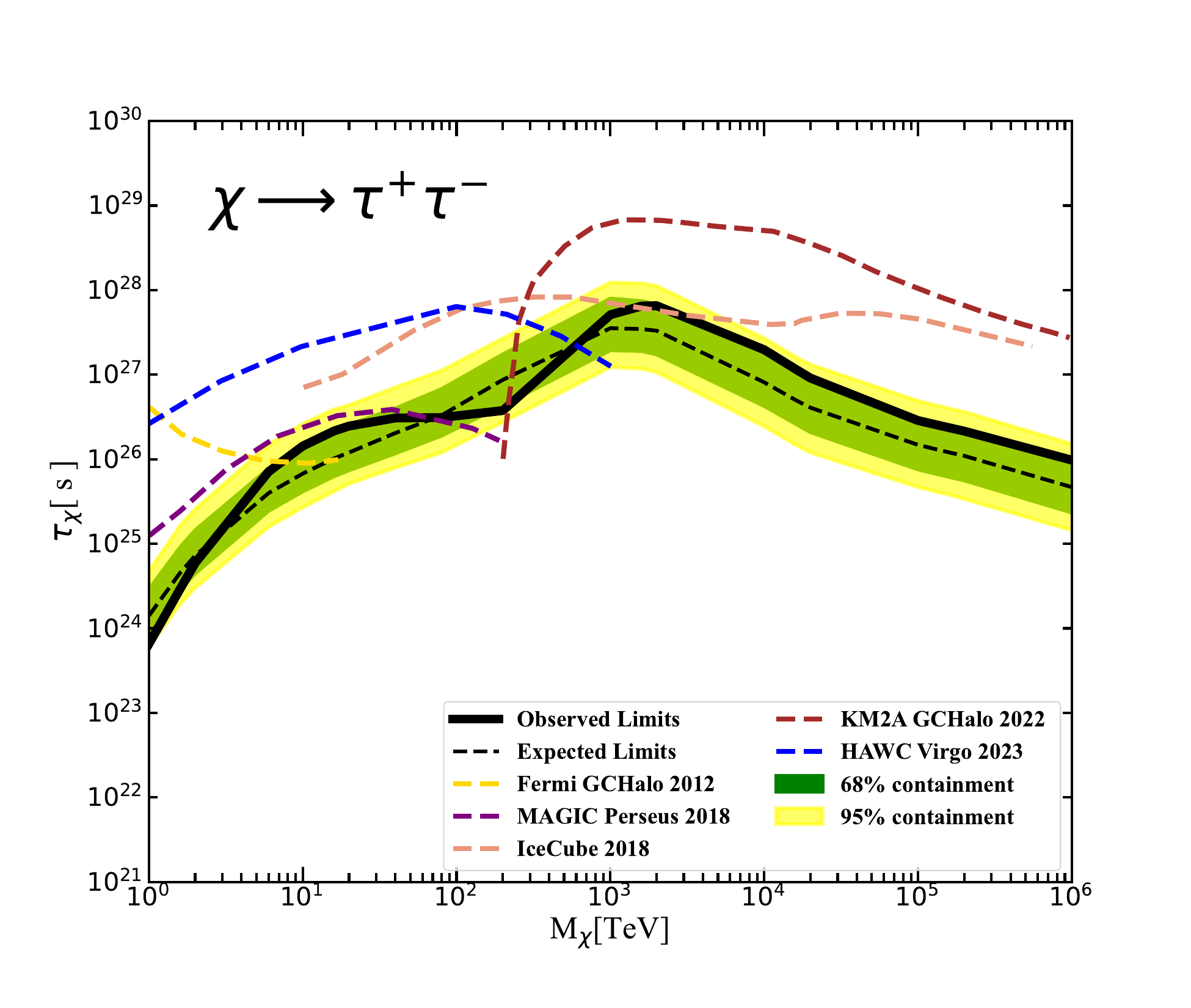}
\caption{The 95$\%$ C.L. lower limits on the DM decay lifetime for  $b\overline{b},\tau^{+}\tau^{-}$ channels. The solid black line represents the LHAASO observed combined limit, and the dashed black line, green band, and yellow band represent the expected combined limits, 1$\sigma$ and 2$\sigma$ uncertainty based on the mock observation. The limits obtained from Fermi-LAT~\cite{Ackermann_2012}, MAGIC~\cite{ACCIARI201838}, IceCube~\cite{Aartsen2018}, LHAASO-KM2A Galactic halo~\cite{KM2A_GCHalo_2022} and HAWC~\cite{PhysRevD.109.043034} with dashed colored line are also shown for comparison.}
\label{expectlimits_decay}
\end{figure*}

\textit{Results}---
We utilize data from 756 days of LHAASO-WCDA and 794 days of LHAASO-KM2A observations to search for DM signals in 16 dSphs around the Milky Way. No significant gamma-ray excess was detected from these dSphs. The statistical significance of DM signals in these dSphs is shown in Sec.~III of Supplemental Material~\cite{Supplement_Materials}. Therefore, 95$\%$ C.L. limits are placed on the DM annihilation cross-section or the DM decay lifetime, as shown in Fig.~\ref{expectlimits_anni} and Fig.~\ref{expectlimits_decay}  respectively.
In Fig.~S8 of Supplemental Material~\cite{Supplement_Materials}, the 95$\%$ C.L. upper limits for $\left \langle \sigma_Av \right \rangle$ from  combined and individual dSphs are presented, assuming a DM mass range from 1 TeV to 1 EeV with a 100$\%$ branching ratio to specific standard model particles. The combined upper limits are dominated by sources with large J-factor, small uncertainties and favorable locations inside the LHAASO FoV, i.e., Ursa Major II, Ursa Minor, Draco, Willman I, Segue 1 and Coma Berenices.

To assess the consistency between the constraints derived from the observed data and the expected limits from pure background, we repeat 1000 mock observations under the null hypothesis, considering the Poisson ﬂuctuation with the measured background. The expected combined limits and the two-sided 68$\%$ and 95$\%$ containment bands for $b\overline{b}$ and $\tau^{+}\tau^{-}$ are shown in Fig.~\ref{expectlimits_anni}. See also Fig.~S10 of the Supplemental Material~\cite{Supplement_Materials} for other channels.  The fact that observed limits are between the expected limit bands indicates that the observational data are consistent with Poisson ﬂuctuation with the background. {The constraints on high-mass DM consistently approach the 68\% boundary of the anticipated limit bands, suggesting a slight overestimation of the background in this study {and thereby a deficit in the number of putative signal events inferred in the ROIs. This
overestimation is likely due to the contribution of faint sources which are below our sensitivity threshold and thus not removed by the mask used in our background estimation; this issue may be more important for high masses due to the lower event rates at high energies.}} Fig.~\ref{expectlimits_anni} also shows the ``Thermal Relic" cross-section~\cite{PhysRevD.86.023506} and the limits from other experiments such as Fermi-LAT~\cite{Fermi_2015}, HAWC~\cite{HAWC_2020}, H.E.S.S~\cite{H.E.S.S_2020}, MAGIC~\cite{MAGIC_2022}, VERITAS~\cite{VERITAS_2023}, and IceCube~\cite{PhysRevD.108.043001}. The observations of dSphs by LHAASO could provide better constraints for DM with a mass heavier than a few hundred TeV.

{In Fig.~S9 of Supplemental Material~\cite{Supplement_Materials}, the 95$\%$ C.L. lower limits for $\tau_{\chi}$ are presented for combined  and individual dSphs analysis.} %\sout{, assuming DM mass range from 1 TeV to 1 EeV with 100$\%$ branching ratio to specific standard model particles.} 
Similar to the DM annihilation results, the limits are mainly driven by Ursa Major II, Ursa Minor, Draco, and Coma Berenices. The expected limits from the same analysis for mock data are shown in Fig.~\ref{expectlimits_decay},  for $b\overline{b}$ and $\tau^{+}\tau^{-}$ final states, with the limits from Fermi-LAT~\cite{Ackermann_2012}, MAGIC~\cite{ACCIARI201838}, IceCube~\cite{Aartsen2018}, LHAASO-KM2A Galactic halo~\cite{KM2A_GCHalo_2022}, and HAWC~\cite{PhysRevD.109.043034}. We also show constraints for $\tau_{\chi}$ from combined dSphs observation and mock data in Fig.~S10 of Supplemental Material~\cite{Supplement_Materials} for other channels.

{In our analysis, we incorporate the J- (D-) factor likelihood into our likelihood analysis, leading to a reduction in the constraints on DM parameters by a factor of 2-6 {(see Fig.~S6 of Supplemental Material~\cite{Supplement_Materials}).} Additionally, we factor in the effects of VHE gamma-ray absorption by the ISRF, resulting in a relaxation of constraints on DM particles with masses exceeding 1000 TeV by approximately 5-10-fold. Moreover, we consider the expected morphology of the DM signal, moving beyond a point-like source approximation. The constraints derived from the extended source analysis based on the DM density profile are consequently diminished by a factor of 1.5-12, particularly in the context of DM decay scenarios, conforming a strong effect of the spatial extension of dSphs on the DM search results \cite{DiMauro:2022hue}. It is important to acknowledge that the J- (D-) factor correction exclusively accounts for the statistical uncertainties in the J- (D-) factors and does not address the systematic uncertainties stemming from the choice of DM profiles and uncertainties with some presumptions about Jeans equation. While factors such as departures from spherical symmetry, velocity anisotropy of the DM halo, the influence of contaminating foreground stars, and variations in the DM profile are considered, the predicted J- (D-) factors and constraints may undergo alterations by a few-fold~\cite{Bonnivard:2014kza,Bonnivard:2015vua,Fermi_2015, Klop:2016lug,Sanders:2016eie,Ichikawa:2016nbi,Ullio:2016kvy,Hayashi:2016kcy}.}

Our results extend for the first time the mass range of the limits on the $\left \langle \sigma_Av \right \rangle$ to 1 EeV with the best constraints above a few hundred TeV. Fermi-LAT~\cite{Fermi_2015}, H.E.S.S~\cite{H.E.S.S_2020}, MAGIC~\cite{MAGIC_2022}, and VERITAS~\cite{VERITAS_2023} exhibit more stringent limits at lower DM masses, since the effective area of LHAASO would decay rapidly at low energy. {We have comparable limits to HAWC~\cite{HAWC_2020} for masses up to several hundred TeV and consistently have better constraints than those from IceCube~\cite{PhysRevD.108.043001} across all mass ranges.}
For the DM decay lifetime, our constraints are weaker than those based on galactic halo data by KM2A \cite{KM2A_GCHalo_2022}, since the D-factor in the selected dSphs is smaller, and the effects of attenuation by pair production are more significant considering the large distances of dSphs from the Earth compared to the galactic halo. {In general, our 
constraints on the DM decay lifetime are also less stringent compared to those determined by HAWC~\cite{PhysRevD.109.043034}, MAGIC~\cite{ACCIARI201838}, Fermi-LAT~\cite{Ackermann_2012}, and IceCube~\cite{Aartsen2018} through the observation of Virgo cluster, Perseus cluster, and the galactic halo with larger D-factors and subdominant effects of attenuation.} However, the combined limits from dSphs, considering the uncertainties of the DM distribution and the spatial extension of the expected signal, could also provide a complementary set of reliable limits.

\textit{Conclusion and Outlook}---
We investigate DM annihilation and decay signals from 16 dSphs within the LHAASO FoV using data collected by WCDA and KM2A. No significant gamma-ray excess is observed from these sources. Consequently, we establish individual and combined constraints on $\left \langle \sigma_Av \right \rangle$ and $\tau_{\chi}$ across five channels ($b\overline{b},t\overline{t},\mu^{+}\mu^{-},\tau^{+}\tau^{-},W^{+}W^{-}$). 

In this analysis, we treat the selected dSphs as extended sources in the 3D likelihood analysis framework to consider the spatial distribution of the DM density within the dSphs. We optimize the size of ROIs and recalculate the J- (D-) factor and their uncertainties corresponding to the ROIs.  To make the analysis more comprehensive and reliable, the absorption effect of ISRF on VHE gamma-ray is considered, and the statistical uncertainty of the J- (D-) factor is incorporated as a nuisance parameter in the likelihood analysis. 

Our results represent the first-ever constraint on $\left \langle \sigma_Av \right \rangle$, extending the mass of DM to 1 EeV. The combined limits are the most stringent constraints for $\left \langle \sigma_Av \right \rangle$ above a few hundred TeV. {Meanwhile, we stress that the impact of spatial extension from dSphs is a necessary condition when deriving DM limits from dSphs with future instruments.} %
As more WCDA and KM2A data will be collected, the development of algorithms to enhance energy and angular resolution for LHAASO , and improvements in kinematics measurements to reduce the uncertainty of the DM density distribution, LHAASO is expected to become more sensitive and to improve its limits in the future.

\textit{Acknowledgements}--- We would like to thank all staff members who work at the LHAASO site above 4400 meters above sea level year-round to maintain the detector and keep the water recycling system, electricity power supply and other components of the experiment operating smoothly. We are grateful to Chengdu Management Committee of Tianfu New Area for the constant financial support for research with LHAASO data. This research work is supported by the following grants: The National Key R$\&$D program of China No.2018YFA0404201, No.2018YFA0404202, No.2018YFA0404203, No.2018YFA0404204,  National Natural Science Foundation of China No.12175248, No.12322302, Department of Science and Technology of Sichuan Province, China No.2021YFSY0030, Project for Young Scientists in Basic Research of Chinese Academy of Sciences No.YSBR-061, the Chinese Academy of Sciences, the Program for Innovative Talents and Entrepreneur in Jiangsu, and in Thailand by the National Science and Technology Development Agency (NSTDA)
and the National Research Council of Thailand (NRCT)
under the High-Potential Research Team Grant Program
(N42A650868).

\bibliography{refs}

%Thailand by the NSRF via the Program Management Unit for Human Resources $\&$ Institutional Development, Research and Innovation (No. B37G660015).

%TC:ignore
\clearpage
\setcounter{figure}{0}
\renewcommand\thefigure{S\arabic{figure}}
\setcounter{table}{0}
\renewcommand\thetable{S\arabic{table}}

\section*{Supplemental Material}
\section{I. Data Analysis with LHAASO}
\label{Appendix_A}

%\subsection{A. The effective observation time}

Table~\ref{tableObsTime} shows the equatorial coordinates and the effective obervation time of LHAASO for target dSphs in this analysis, where the $\rm T_{WCDA}$ and $\rm T_{KM2A}$ represent the effective obervation time of LHAASO-WCDA and LHAASO-KM2A respectively.

\begin{table*}[htbp]
    \caption{The associated the right ascension, the declination, and the effective observation time (day) of WCDA and KM2A for 16 dSphs within the LHAASO FoV.}
    \label{tableObsTime}
            
    \centering 
	\renewcommand\tabcolsep{11.0pt}
    \begin{tabular}{c c c c c}
        \botrule
            Name  & Ra[deg] & Dec[deg] &$ \rm T_{WCDA}$[day] &$ \rm T_{KM2A}$ [day] \\
            % \botrule
            \hline
Draco & 260.05 & 57.92 & 690.37 & 730.76 \\
Ursa Minor & 227.28 & 67.23 & 701.59 & 739.53 \\
Ursa Major I & 158.71 & 51.92 & 692.51 & 742.17 \\
Ursa Major II & 132.87 & 63.13 & 688.88 & 740.83 \\
Bootes 1 & 210.02 & 14.50 & 699.85 & 742.01 \\
Canes Venatici I & 202.02 & 33.56 & 699.01 & 743.28 \\
Coma Berenices & 186.74 & 23.90 & 696.88 & 744.34 \\
Leo I & 152.12 & 12.30 & 692.19 & 743.02 \\
Segue 1 & 151.77 & 16.08 & 691.72 & 743.06 \\
Sextans & 153.26 & -1.61 & 693.36 & 743.60 \\
Canes Venatici II & 194.29 & 34.32 & 698.15 & 743.81 \\
Hercules & 247.76 & 12.79 & 699.04 & 737.33 \\
Leo II & 168.37 & 22.15 & 693.95 & 744.18 \\
Willman I & 162.34 & 51.05 & 693.45 & 742.21 \\
Aquarius 2 & 338.48 & -9.33 & 666.84 & 728.11 \\
Leo T & 143.72 & 17.05 & 690.84 & 742.51 \\
        \botrule
    \end{tabular}
\end{table*}

\begin{figure*}[!htb]

	\includegraphics[width=0.7\textwidth]{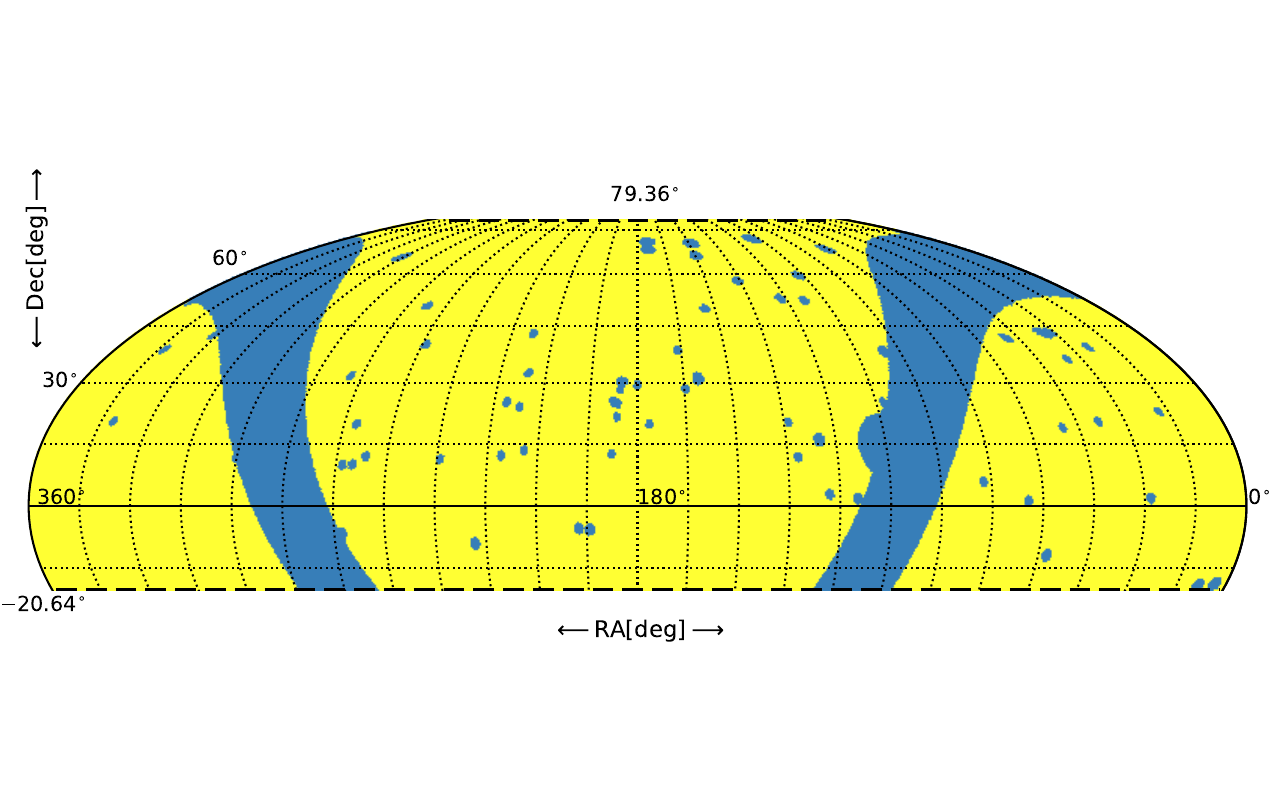}
	
	\caption{The masked sky map in the equatorial coordinate system.}
% \caption{The masked sky map in the \textcolor{blue}{galactic coordinate system}.}
	\label{mask_regions}
\end{figure*}

%\subsection{B. Mask Regions for Background Estimation}
Fig~\ref{mask_regions} displays the masked sky map during background estimation in the equatorial coordinate system. In this work, we mask the Galactic disk region ($|\rm{b}| < 10^{\circ}$), known sources given by TeVCat~\cite{Wakely:2007qpa} and first LHAASO catalog~\cite{LHAASO:2023rpg} with exclusion radius $R_{excl}=n\cdot\sqrt{\sigma_{ext}^2+\sigma_{psf}^2}$, where $n=2.5$ is a constant factor, $\sigma_{ext}$ and $\sigma_{psf}$ denote the sources extension and the PSF of instruments respectively, as in Ref.~\cite{LHAASO:2023gne}.
%And the 23 dSphs in the LHAASO's field of view are also masked with a radius of $10^{\circ}$.

%\subsection{C. Detector Response}
Detector response for gamma-ray is crucial to relate the DM flux to the number of events from the dSphs.
We investigate the detector response of LHAASO-WCDA and LHAASO-KM2A to detect gamma-ray as a function of the primary energy and the incidence direction from same detector simulation procedure in performance paper\cite{Aharonian_2021_a,Aharonian_2021}.
Following Ref.~\cite{Huang:2021yfu}, the expected events (${N_{i,j,k}^{DM}}$) of gamma-ray from DM annihilation or decay in $i$-th energy estimator bin and $ j$-th pixel of 2D sky map for $ k$-th dSph in the observation time is calculated by
\begin{equation}
                {N_{i,j,k}^{DM}=\int_{i} {\rm d}\eta\int_{j}{\rm d}\Omega(\hat{p'})\int {\rm d}t R(\eta,\hat{p'},t) },
            \end{equation}
where the $\rm \eta$ is the energy estimator, which represented as the number of triggered PMTs ($\rm N_{hit}$) for WCDA, and represented as reconstruction energy for KM2A. The $\hat{p'}$ represents the reconstructed direction according to the direction reconstruction algorithm and $ t$ denotes the observation time. As for the R, it can be described by the following formation, 
\begin{equation}
                {R(\eta,\hat{p'},t)=\int {\rm d}E\int_{\hat{p}}{\rm d}\Omega\frac{{\rm d}F}{{\rm d}E{\rm d}\Omega {\rm d}t} A(\hat{p},E) P(\hat{p'},E;\hat{p})M(\eta,\hat{p};E)}.
                \label{M}
            \end{equation}
Here, the $E$ and $\hat{p}$ are the primary energy and  primary direction for a gamma-ray event. The flux term is expected differential gamma-ray flux from DM annihilation or decay, which is described as Eq.1  and Eq.2 of the Letter. These three factors are computed for once LHAASO observation, effective collection area 
($ A(\hat{p},E)$), point-spread-function ($ P(\hat{p'},E;\hat{p})$) and energy estimator transfer matrix ($ M(\eta,\hat{p};E)$). According to LHAASO official simulation data\cite{Aharonian_2021_a,Aharonian_2021}, the effective area is the product of reference area and efficiency from event trigger, reconstruction and selection for gamma-ray with primary energy $E$ and primary direction $\hat{p}$. The point spread function and the energy estimator transfer matrix are the probability distributions of the reconstructed direction $\hat{p'}$ and the observed energy estimator with primary energy $E$ and primary direction $\hat{p}$ respectively. With the expected events of gamma-ray from DM, ${N_{i,j,k}^{DM}}$, and estimated background events, we could get the expected numbers of gamma-ray, $N_{i,j,k}^{exp}$, in the $ i$-th energy estimator ($N_{hit}$ for WCDA and reconstructed energy for KM2A) bin and the $ j$-th pixel on the 2D sky map of the $ k$-th dSph.

\section{II. Calculation for J- (D-) factor and density profile}

{The observation targets refered in this work are selected from Ref.~\cite{Pace_2018}.  Among the 44 dSphs objects studied in Ref.~\cite{Pace_2018}, 
we exclude the 6 dSphs bound to Andromeda Galaxy (M31) and 15 dSphs that are not in LHAASO's FoV, where the declination of dSphs satisfies the range $-20.64^{\circ}$ to $79.36^{\circ}$. 
%$-20.64^{\circ}$ to $79.36^{\circ}$).
Subsequently, from the remaining 23 dSphs, we select 16 dSphs as our primary observation targets. The coordinates of these selected dSphs are detailed in Table \ref{tableObsTime}. Notably, apart from their advantageous locations, these chosen dSphs exhibit tighter constraints compared to the unselected dSphs.
{The seven dSphs within LHAASO's FoV which the authors of Ref.~\cite{Pace_2018} indicate should be treated with caution because the MCMC chains include tails or provide only upper limits in posterior distributions (Draco II, Leo IV, Leo V, Pisces II, Pegasus III, Segue 2 and Triangulum II), are not selected for the study.}}

In Eq.~1 and Eq.~2 of the Letter, $ {\rm d}J/{\rm d}\Omega$ and $ {\rm d}D/{\rm d}\Omega$ are defined as, 

\begin{equation}
            \begin{aligned}
            { \frac{{\rm d}J}{{\rm d}\Omega}=\int\rho^{2}_{DM}(r){\rm d}l}, 
            \label{con:jfactor}
            \end{aligned}
            \end{equation}
\begin{equation}
            \begin{aligned}
            { \frac{{\rm d}D}{{\rm d}\Omega}=\int\rho_{DM}(r){\rm d}l}, 
            \label{con:dfactor}
           \end{aligned}
\end{equation}
where the  $\rm {\rho_{DM}(r)}$ refers to the DM density profile in the dSphs, and $l$ represents the distance from a point on the line-of-sight (L.o.S) to the Earth. The relationship between $r$ and $l$ is described by $ r^{2}=l^{2}+d^{2}-2ldcos\theta$. Here, $d$ represents the distance between the dSph center and the Earth, while $\theta$ denotes the angle between the L.o.S and the direction of the dSph center. In this work, Navarro–Frenk–White (NFW)\cite{DMProfile_1997} model is adopted as
\begin{equation}
	\label{rho}
	\begin{aligned}
		{ \rho_{DM}(r)=\frac{\rho_{s}}{(r/r_s)(1+r/r_s)^{2}}},
	\end{aligned}
\end{equation}     
where $\rho_s$ is the scale density and $r_{s}$ is the scale radius. The upper limits of integration of $l$ are conservatively estimated with the tidal radius ($r_t$) of dSphs \cite{Pace_2018}, as
\begin{equation}
	\begin{aligned}
		{ l_{\pm}=d{\rm cos}(\theta)\pm \sqrt{ r_t^2-d^2{\rm sin}^{2}(\theta)}}.
	\end{aligned}
\end{equation}

\begin{figure*}[htbp]
\includegraphics[height=0.3\textwidth,width=0.4\textwidth]{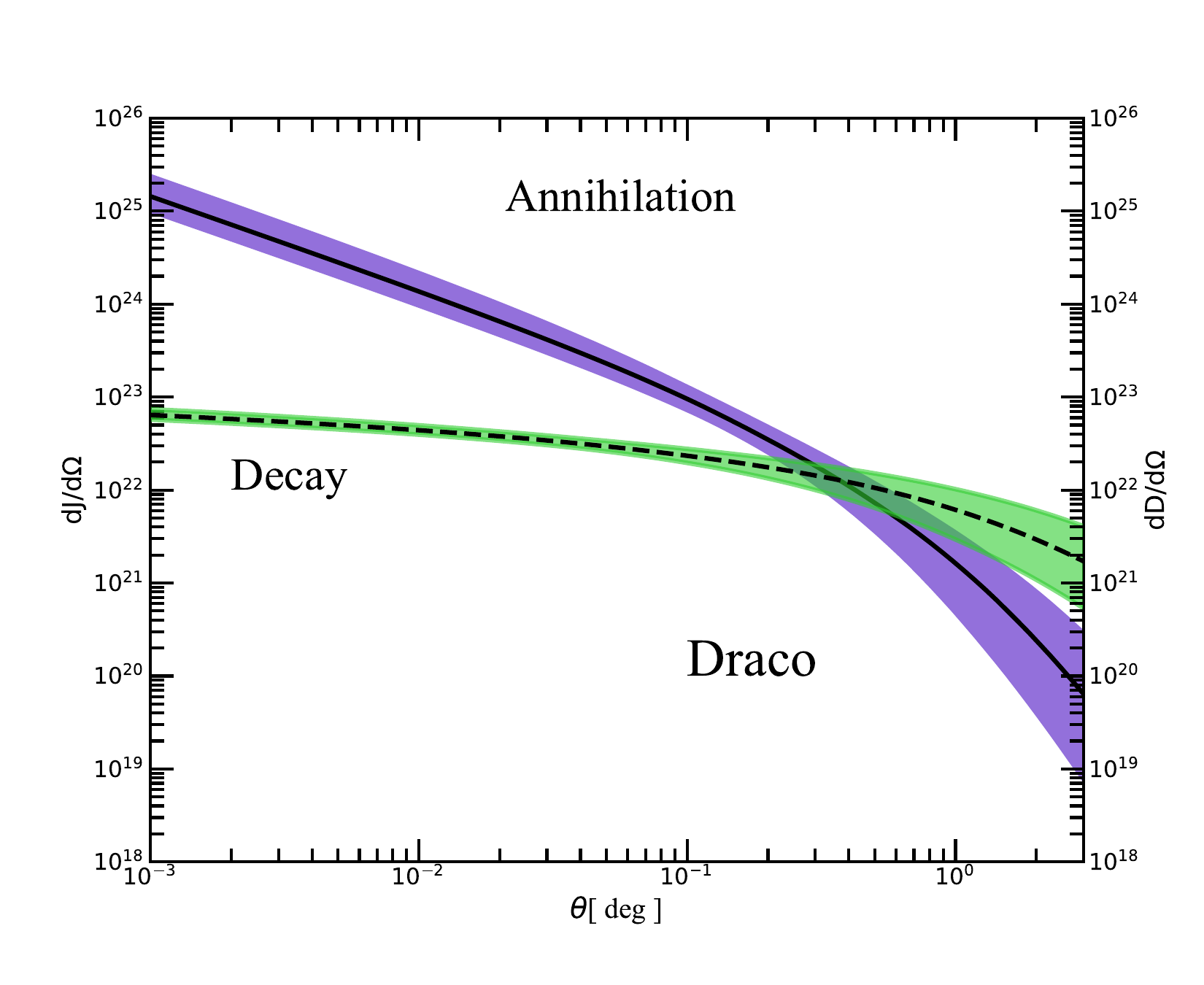}
\includegraphics[height=0.3\textwidth,width=0.4\textwidth]{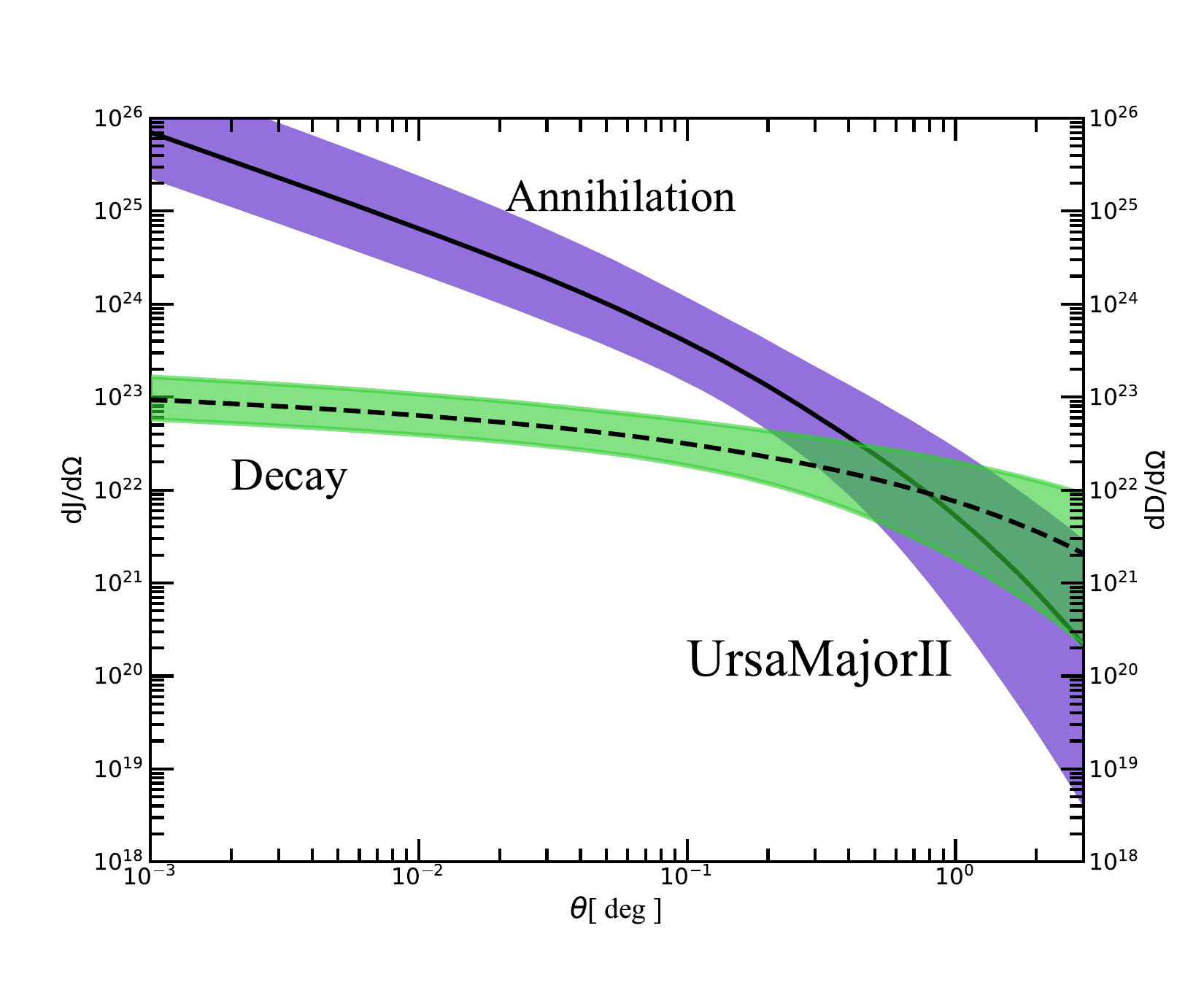}
\caption{Expected differential J- profile (purple) and D- profile (green)
for Draco and Ursa Major II. At each angle the solid and dashed lines show the median
proﬁles and the shaded band corresponds to the $\pm1 \sigma$ distribution as derived
in MCMC chains provided by Ref~\cite{Pace_2018}.}
\label{profile}
\end{figure*}

{The parameters (i.e. $d$, $r_{t}$, $\rho_{s}$, $r_{s}$) determine the calculation of the density profile and the J- (D-) factor of dSphs. The Ref~\cite{Pace_2018} performed a Jeans analysis on the target dSphs, whose DM profiles follow the NFW profile with these parameters mentioned above. We use their public available MCMC chains to calculate the distribution of density profile and J- (D-) for each target dSph.
Fig~ \ref{profile} shows the constraints on the differential J- (D-) profile for two dSphs: the classical dwarf Draco and the ultra-faint dwarf Ursa Major II. The envelopes show the $\pm1 \sigma$ and median values of the differential J- (D-) profile as a function of the angular separation from the center of the dSphs. }

%\begin{figure}[htbp]
%\includegraphics[height=0.3\textwidth,width=0.4\textwidth]{sampleJD_draco1.pdf}
%\includegraphics[height=0.3\textwidth,width=0.4\textwidth]{sampleJD_ursamajor2.pdf}
%\caption{Expected J- factor (purple) and D- factor (green)
%for Draco and UrsaMajorII. At each angle the solid and dashed lines show the medianproﬁles and the shaded band corresponds to the $\pm1 \sigma$ distribution as derivedin MCMC chains provided by Ref~\cite{Pace_2018}.}
%\label{profile}
%\end{figure}

\begin{table}[!htbp]%
            %\begin{center}
            \tiny
            \begin{minipage}{0.48\textwidth}
            \caption{The corresponding density profile parameters of J- factor for each dSphs ~\cite{Pace_2018}.}
            \label{tabJ_D}%
            
            \centering
            \resizebox{\textwidth}{!}{
            \begin{tabular}{@{}ccccc@{}}
            \botrule
            Source & $d/\rm{kpc}$&  ${r_{t}}/\rm{kpc}$ &${\rm log_{10}(}\rho_{s}/M_{\odot}{\rm kpc^{-3})}$& ${\rm log_{10}(}r_{s}/{\rm kpc})$ \\
            % \botrule
            \hline
Draco & 67.40 & 9.07 & 7.45 & 0.17\\
Ursa Minor & 73.29 & 6.92 & 7.91 & -0.18\\
Ursa Major I & 95.58 & 8.34 & 7.83 & -0.19\\
Ursa Major II &35.12 &9.35 &6.79 &0.72\\
Bootes 1 & 65.69 & 9.06 & 6.62 & 0.58\\
Canes Venatici I & 211.09 & 19.82 & 7.20 & 0.14\\
Coma Berenices & 42.31 & 5.14 & 7.57 & 0.06\\
Leo I & 266.15 & 23.25 & 7.72 & -0.09\\
Segue 1 & 23.88 & 3.41 & 6.96 & 0.39\\
Sextans & 89.63 & 11.88 & 6.32 & 0.65\\
Canes Venatici II & 166.26 & 29.91 & 6.50 & 0.70\\
Hercules & 138.16 & 22.21 & 6.17 & 0.82\\
Leo II & 242.44 & 19.15 & 7.95 & -0.23\\
Willman I & 44.46 & 8.07 & 7.81 & 0.10\\
Aquarius 2 & 107.50 & 24.46 & 6.59 & 0.78\\
Leo T & 380.69 & 25.00 & 6.69 & 0.74\\
\botrule
\end{tabular}
}
\end{minipage}
\end{table}

\begin{figure*}[htbp]
\includegraphics[height=0.31\textwidth,width=0.4\textwidth]{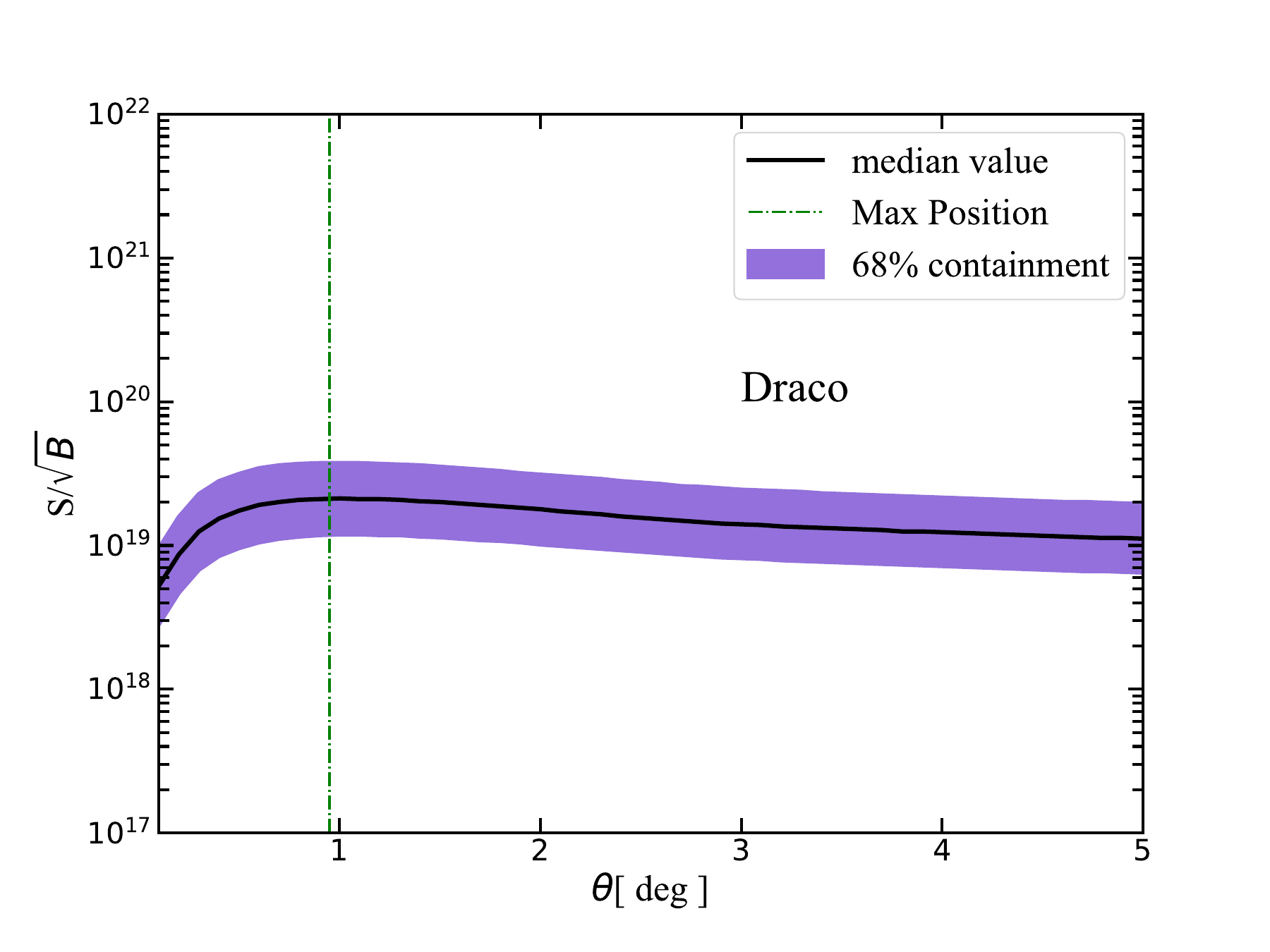}
\includegraphics[height=0.31\textwidth,width=0.4\textwidth]{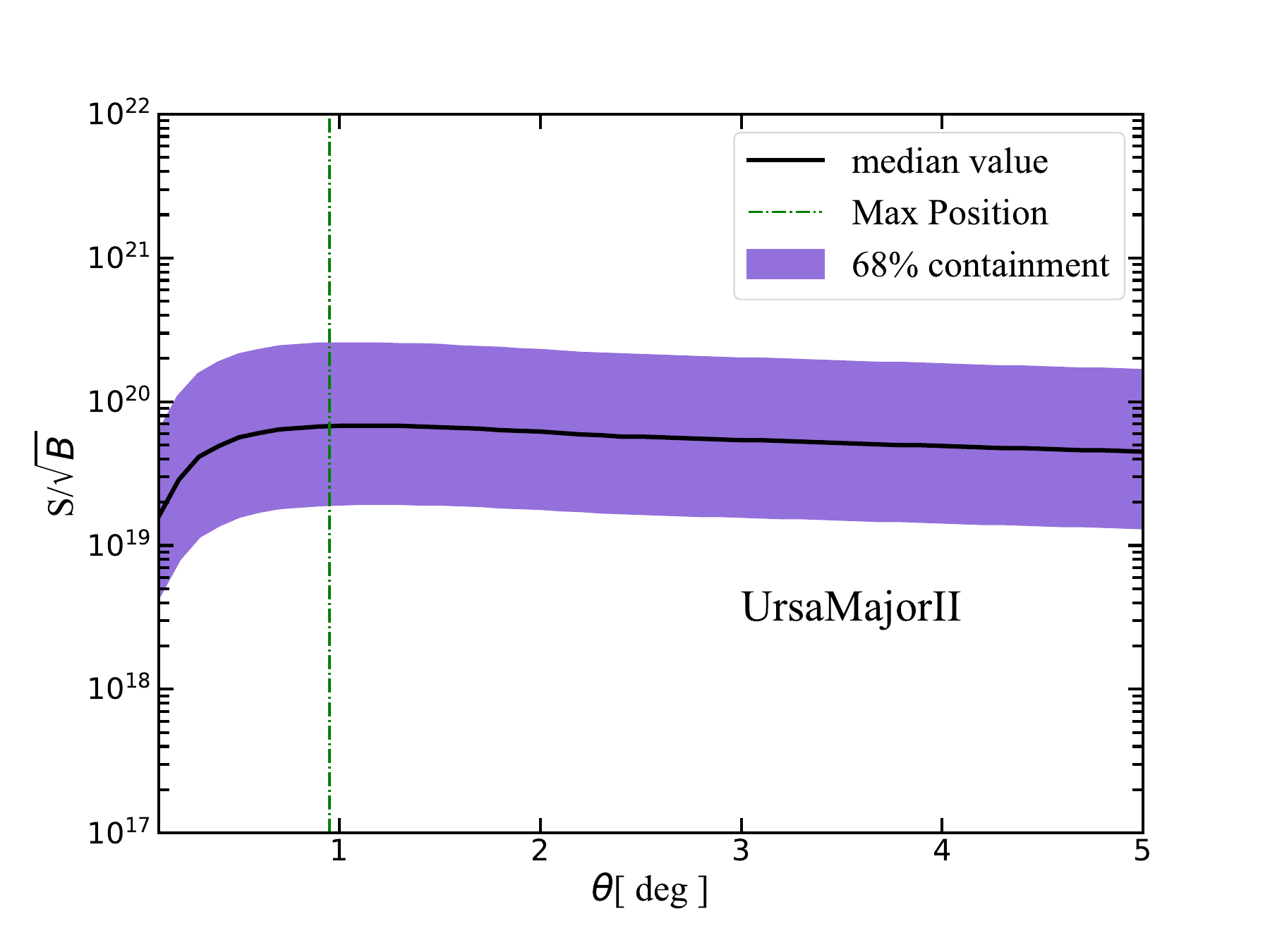}
\caption{The DM annihilation signal-to-noise ratio as a function of the angle for the Draco (left) and Ursa Major II (right). The black solid line represents the median curve of signal-to-noise ratio, and the shaded band corresponds to the $\pm1 \sigma$ distribution. The maximum value of the median curve is represented by the green dashed line, corresponding to the optimal  region.}
\label{fig:S_B_PDF}
\end{figure*}

{In order to maximize the sensitivity of the signal search, we use $\mathcal S/\sqrt{\mathcal B}$ as an indicator to find the optimal region of interest (ROI)~\cite{Li:1983fv}, where $\mathcal S$ is the 2D signal map, which use the J- (D-) profile of each dSphs convolved with 2D point spread function (PSF) at first KM2A energy estimator bin and multiply by the solid angle in each pixel. $\mathcal B$ is background distribution estimated according to the background estimation method mentioned in data analysis. We sum all signal within $\theta$ to get $\mathcal S$ and sum all background within $\theta$ to get $\mathcal B$, and we vary $\theta$ to maximize $\mathcal S/\sqrt{\mathcal B}$. We obtain the variation of the signal-to-noise ratio($\mathcal S/\sqrt{\mathcal B}$) with the angle $\theta$, and select the $\theta$ with the maximum median signal-to-noise ratio as the ROI size. Fig~\ref{fig:S_B_PDF} shows the DM annihilation signal-to-noise ratio as a function of the $\theta$ for the Draco and Ursa Major II, in which the black solid line represents the median curve of signal-to-noise ratio, and the maximum value of the median is represented by the green dashed line, corresponding to the optimal  region. The optimal regions for some dSphs are smaller than the PSF of KM2A at first KM2A energy estimator bin, in which case we fixed the selected ROI as 1.58$\sigma$ ~\cite{Alexandreas:1992ek}, and use the parameters corresponding to the median of the J-(D-) factor at the optimal ROI as the parameters of the dark matter distribution model, as shown in Table \ref{tabJ_D}. The ROI sizes and J- (D-) factors of the 16 dSphs are shown in Table I in the Letter.}

\section{III. Statistical Significance of DM Annihilation and Decay Signals}
\label{Appendix_B}

\begin{figure*}[htbp]
\centering
	\includegraphics[width=0.45\textwidth]{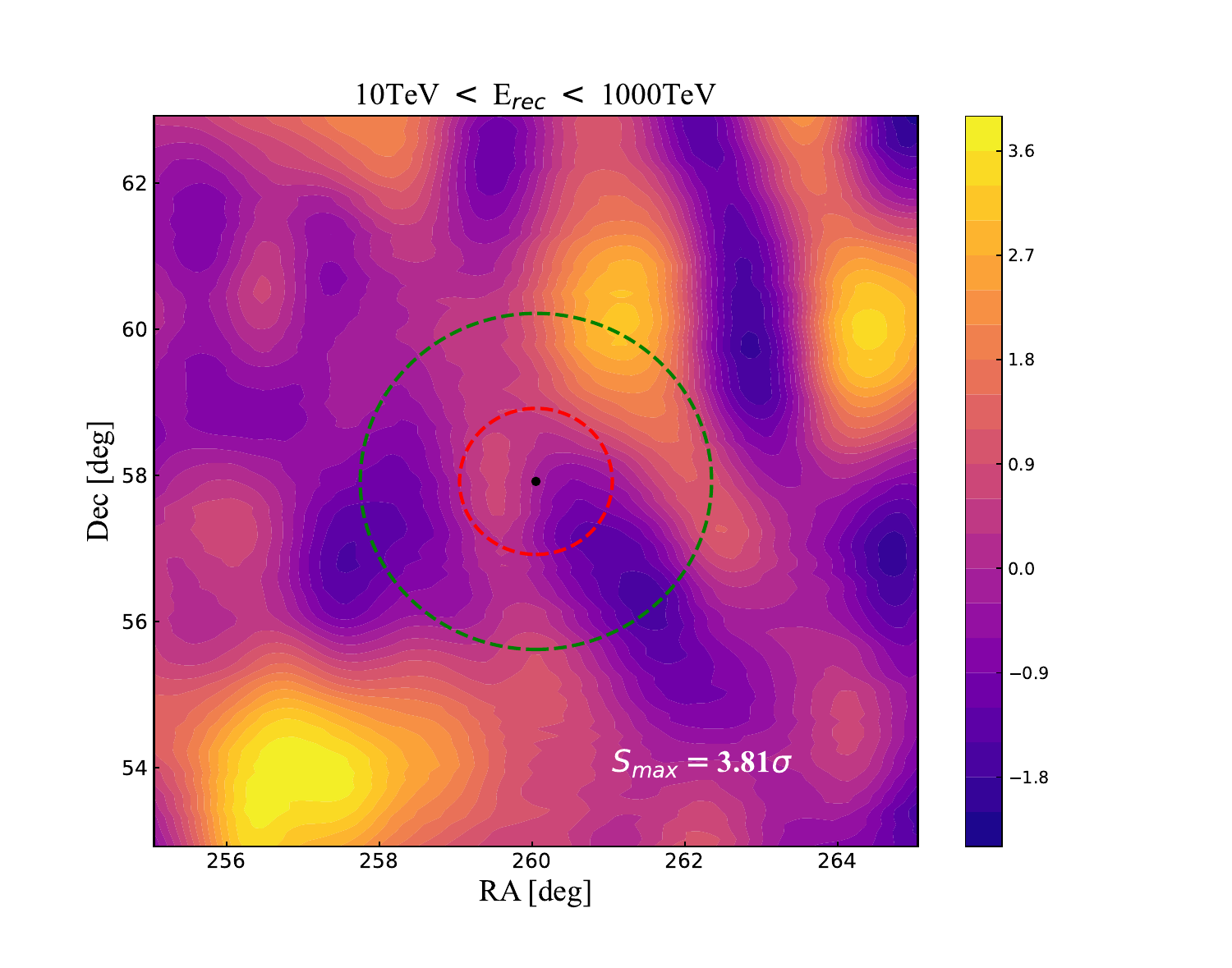}
	\includegraphics[width=0.45\textwidth]{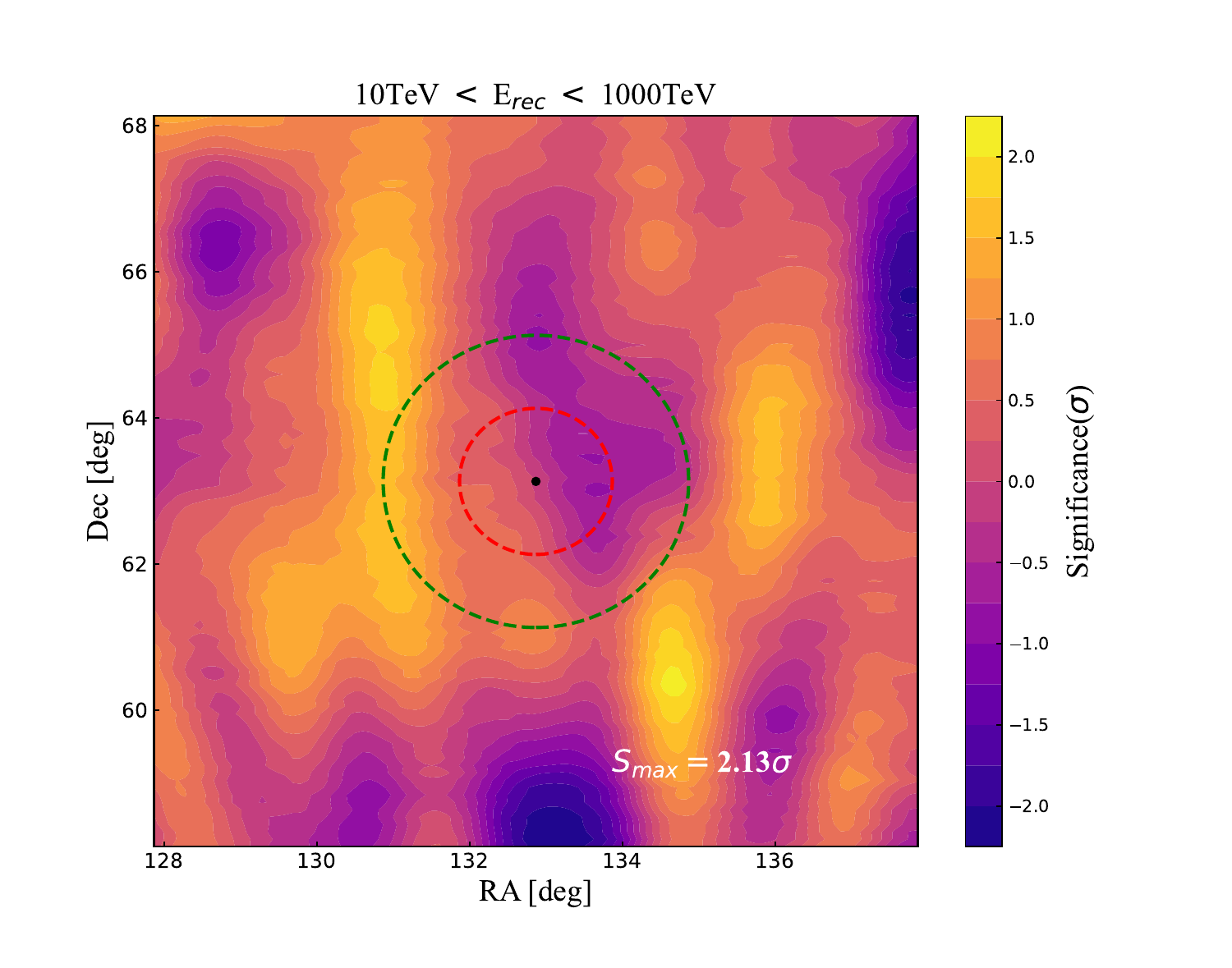}
	\includegraphics[width=0.45\textwidth]{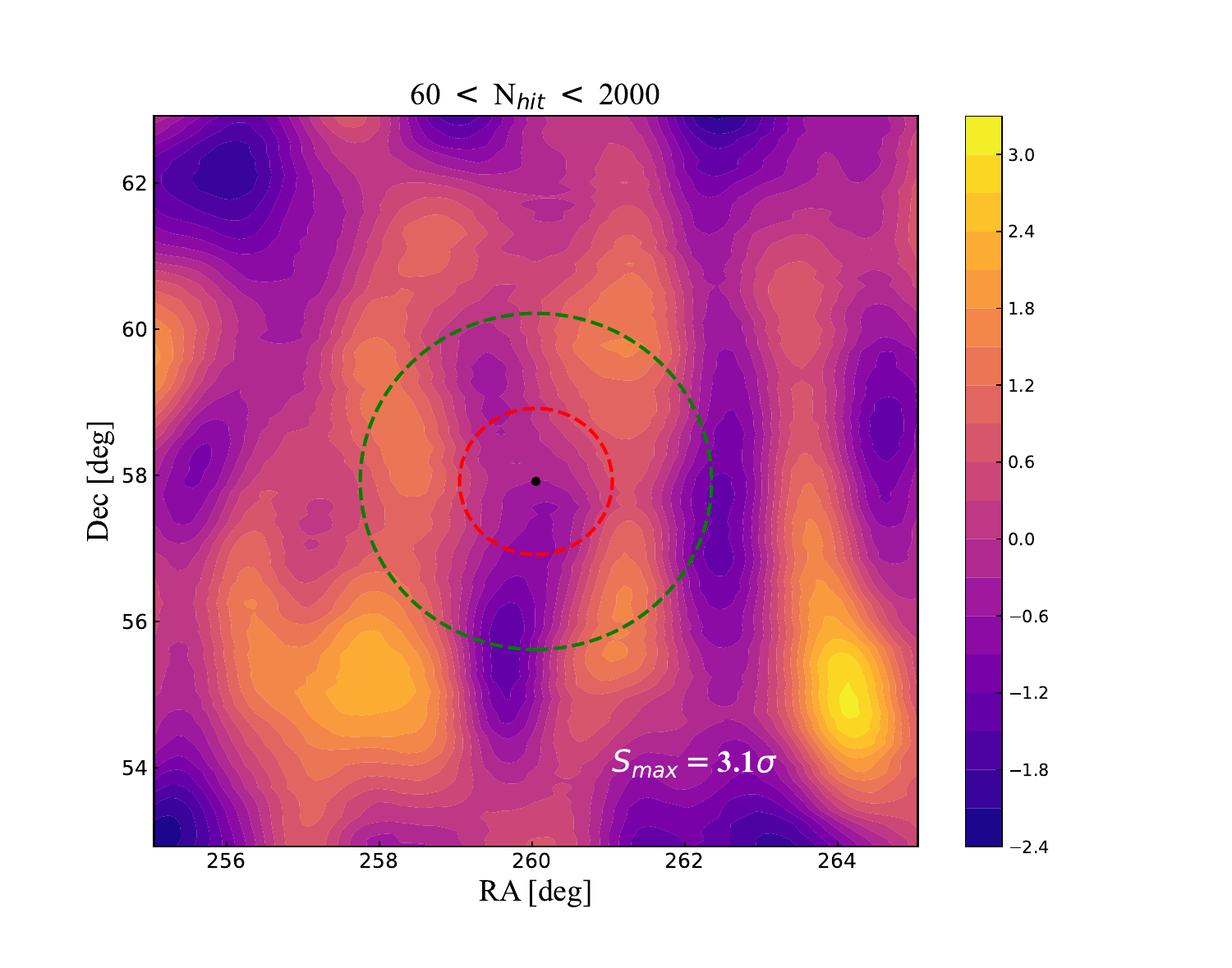}
	\includegraphics[width=0.45\textwidth]{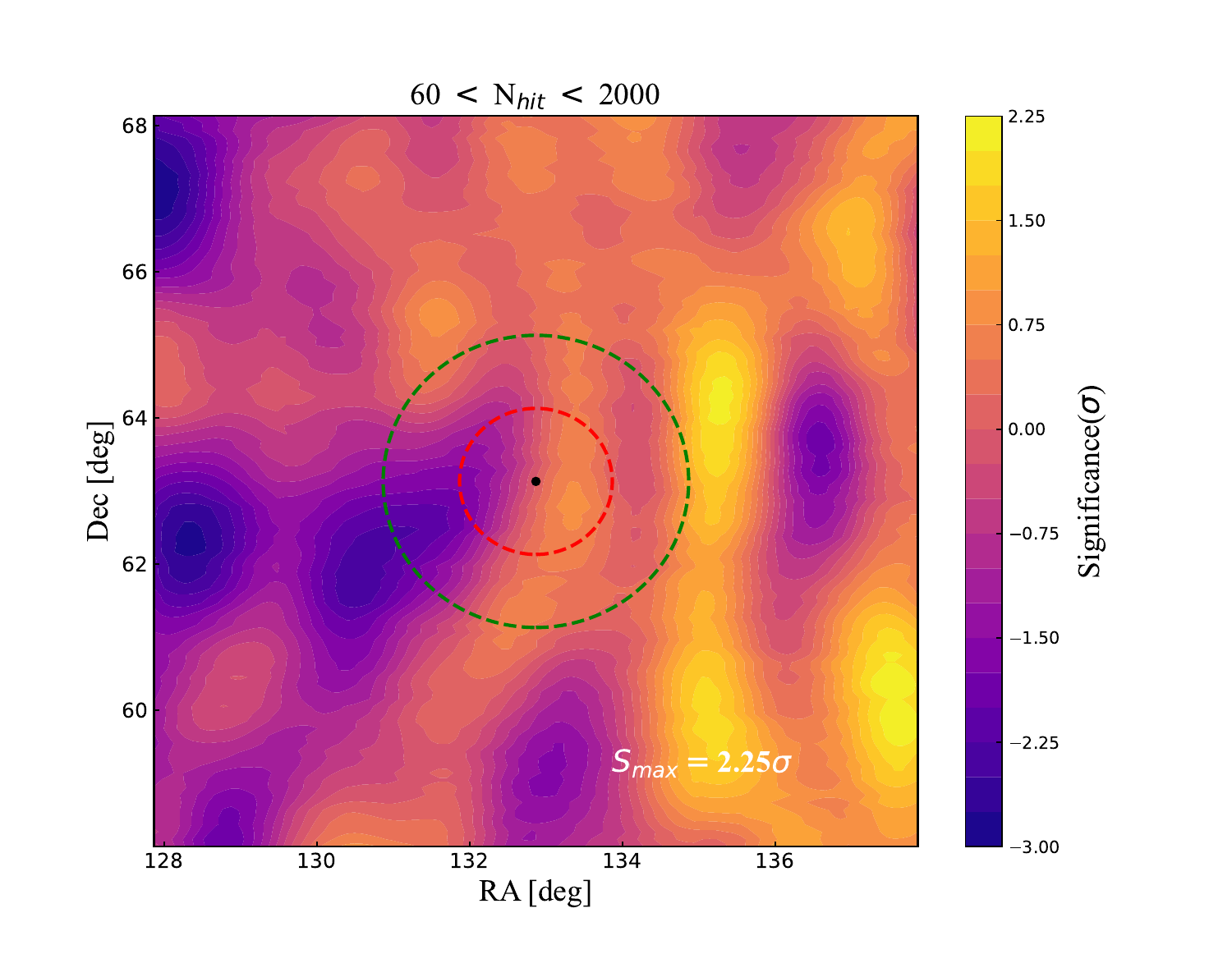}
	
	\caption{The significance map of Draco (left) and Ursa Major II (right) of the energy of 10-1000 TeV in the KM2A data (top) and $N_{hit}$ of 60-2000 in the WCDA data (bottom).}
	\label{TS_map_KM2A_LIMA}
\end{figure*}

{The significance map of two dSphs, the classical dwarf Draco and the ultra-faint dwarf Ursa Major II, are calculated by using the Li-Ma formula \cite{Li:1983fv} with KM2A and WCDA data, as shown in Fig~\ref{TS_map_KM2A_LIMA}. The dashed green and red lines in the figures indicate the ROI regions for DM decay and annihilation, respectively. }

\begin{figure*}[htbp]
                	
                \includegraphics[height=0.31\textwidth,width=0.4\textwidth]{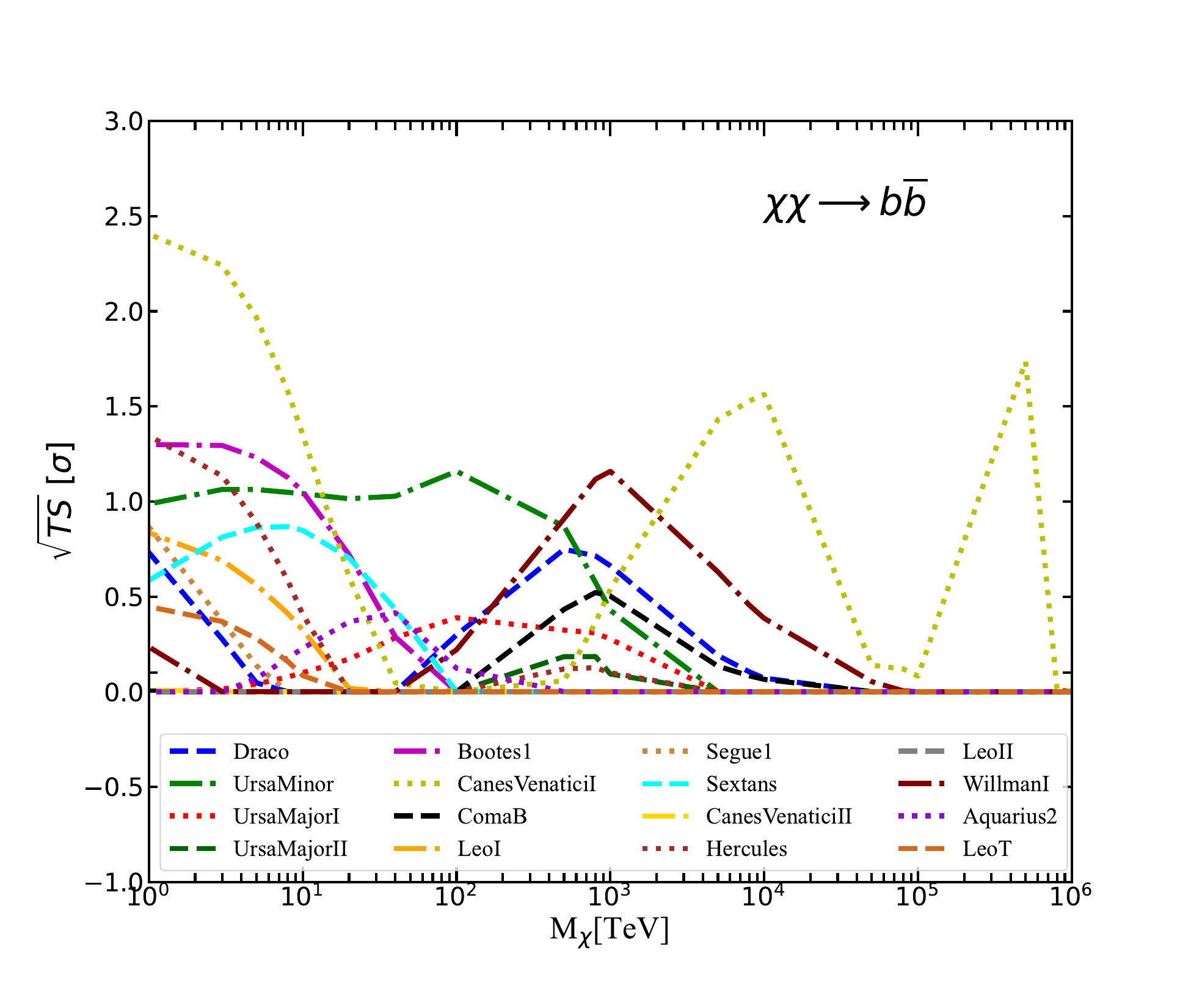}
                \includegraphics[height=0.31\textwidth,width=0.4\textwidth]{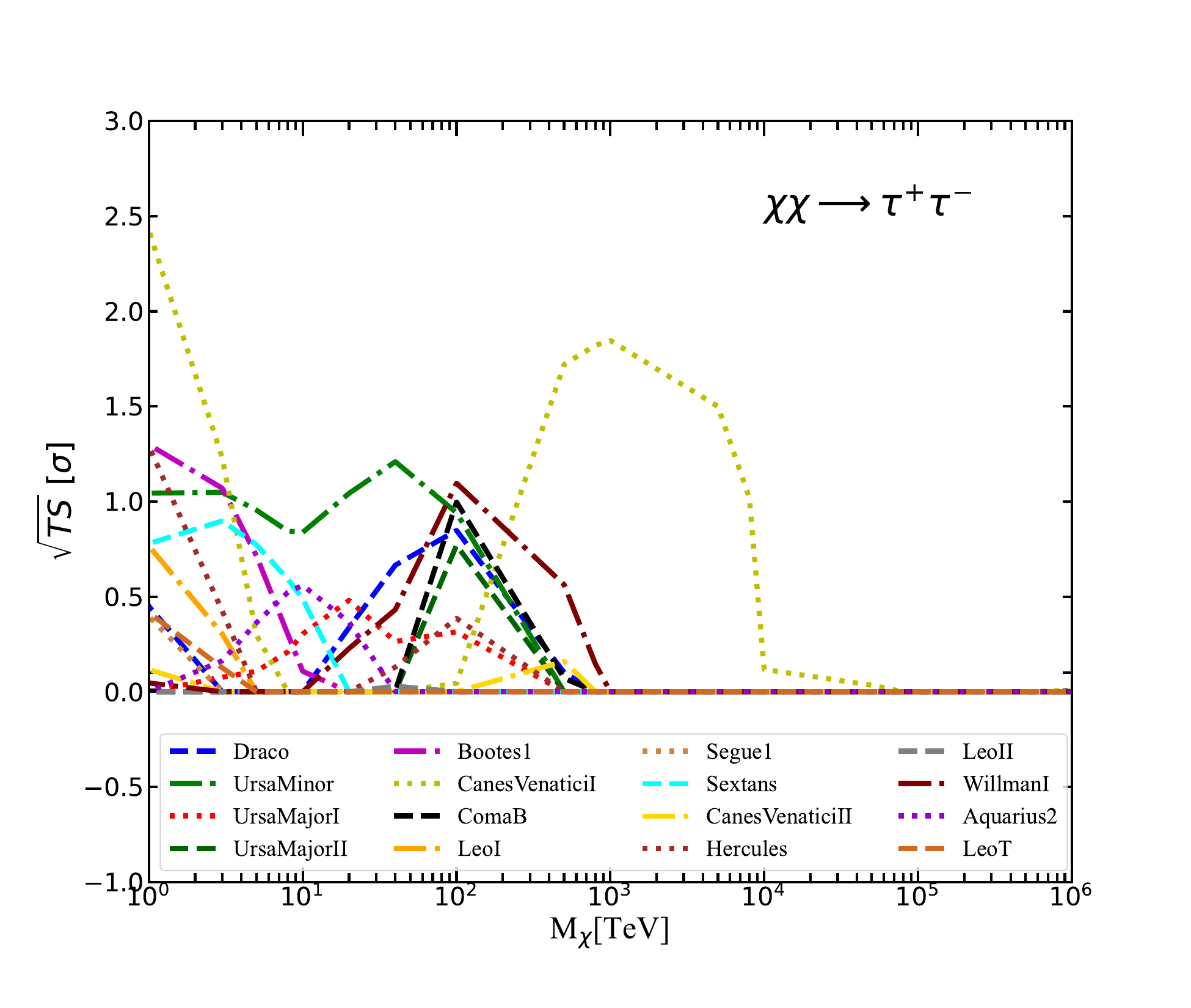}
                \includegraphics[height=0.31\textwidth,width=0.4\textwidth]{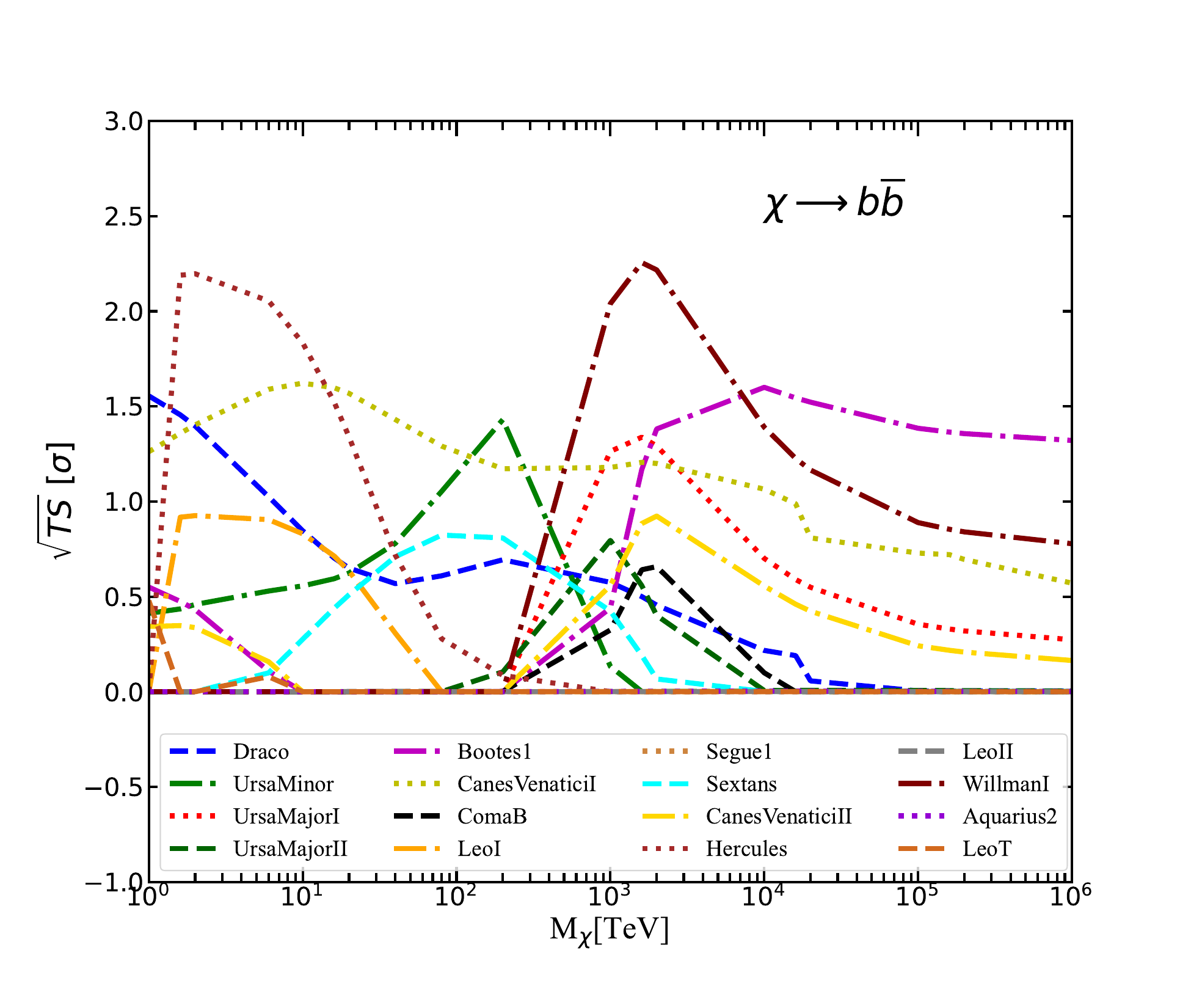}
                \includegraphics[height=0.31\textwidth,width=0.4\textwidth]{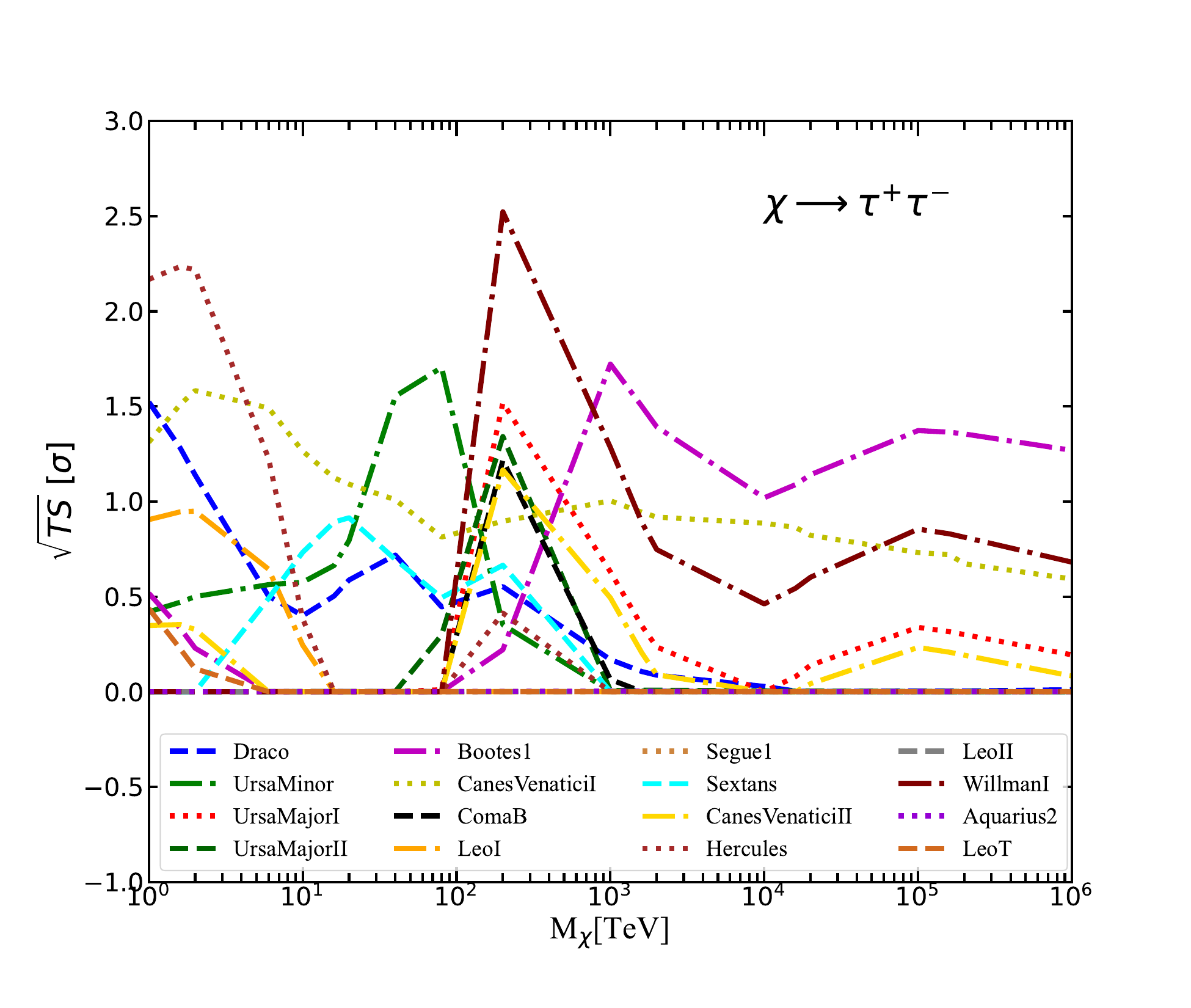}
                 \caption{ LHAASO-measured signiﬁcance of the dark matter annihilation (top panel) and decay (bottom panel) signal in two channels, i.e., ${b\overline{b}}$ and ${\tau^{+}\tau^{-}}$, for the 16 individual dSphs. The color lines show the signal signiﬁcance for each dSphs as a function of dark matter particle mass. }
                
                \label{TS}
            \end{figure*}

Fig \ref{TS} shows the signal signiﬁcance as a function of dark matter particle mass in the two channels, i.e., $b\overline{b}$ and $\tau^{+}\tau^{-}$. There is no significant gamma-ray signal excess from the dark matter annihilation or decay process in this analysis, with the highest significance approaching 2.4 $\sigma$. 

%\FloatBarrier
% \newpage
\nopagebreak[4]

\section{IV. Other supplementary results}
\label{Appendix_C}
\begin{figure*}[htbp]
                	\centering
                	\includegraphics[width=0.4\textwidth]{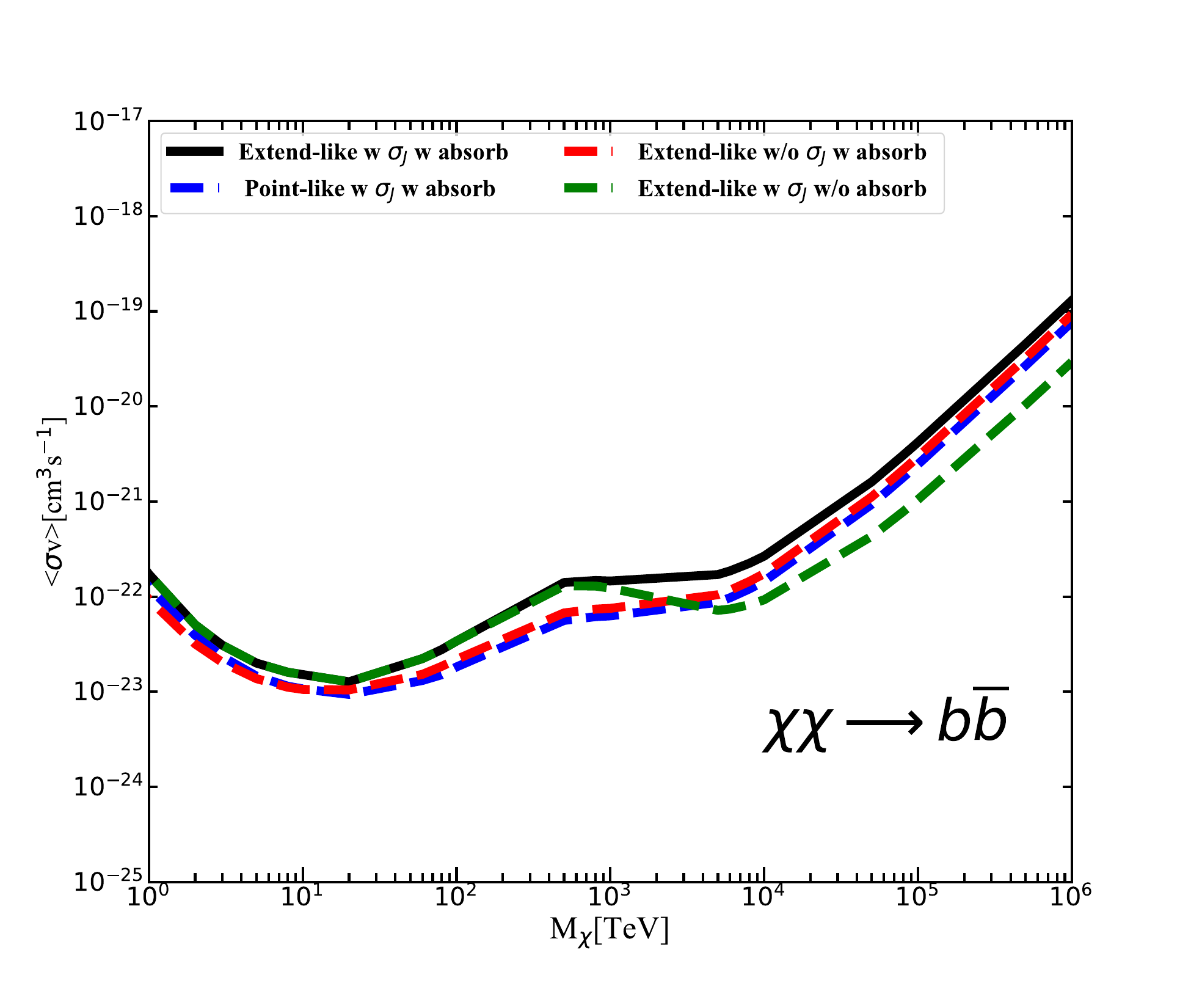}
                	\includegraphics[width=0.4\textwidth]{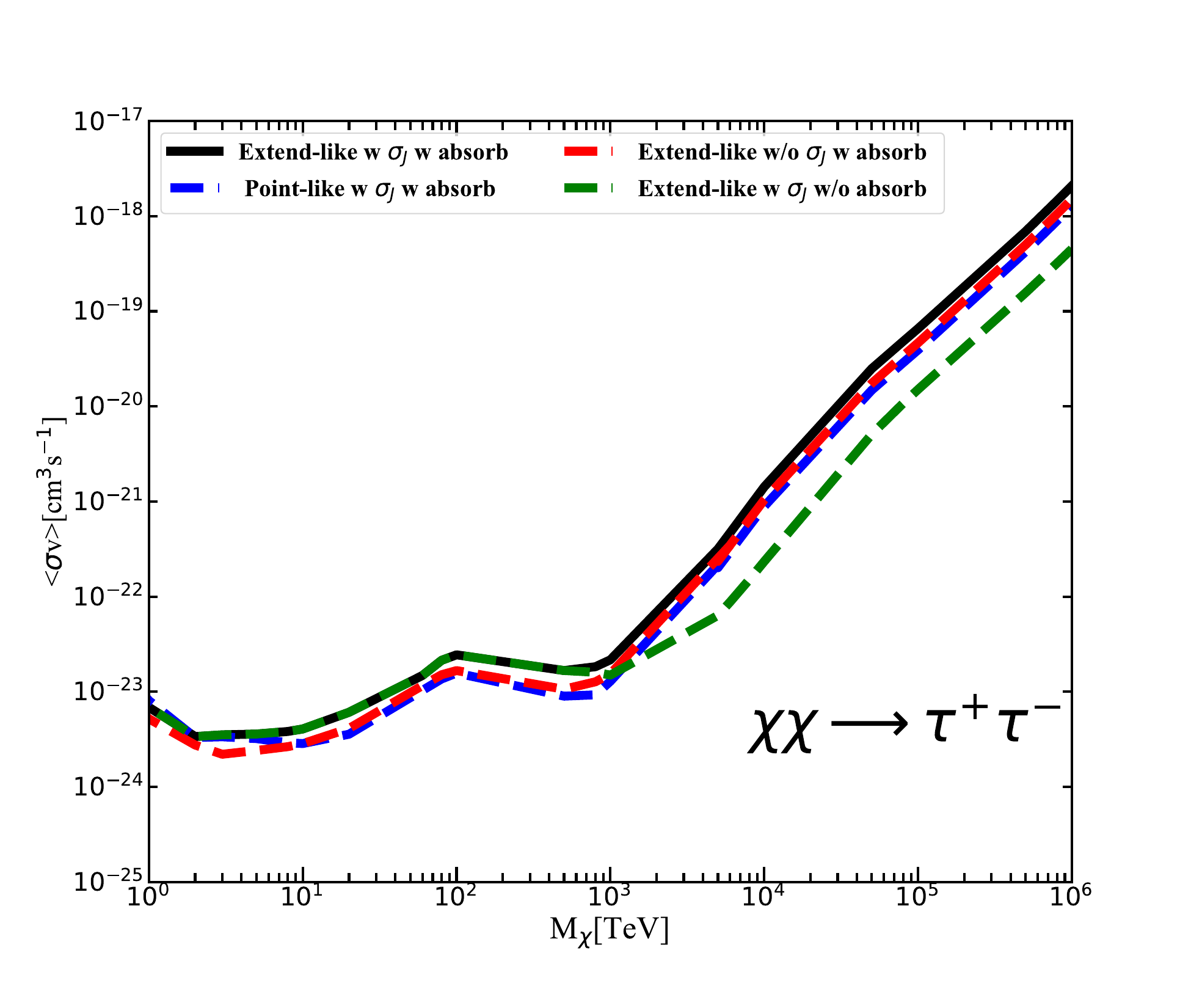}
                	\includegraphics[width=0.4\textwidth]{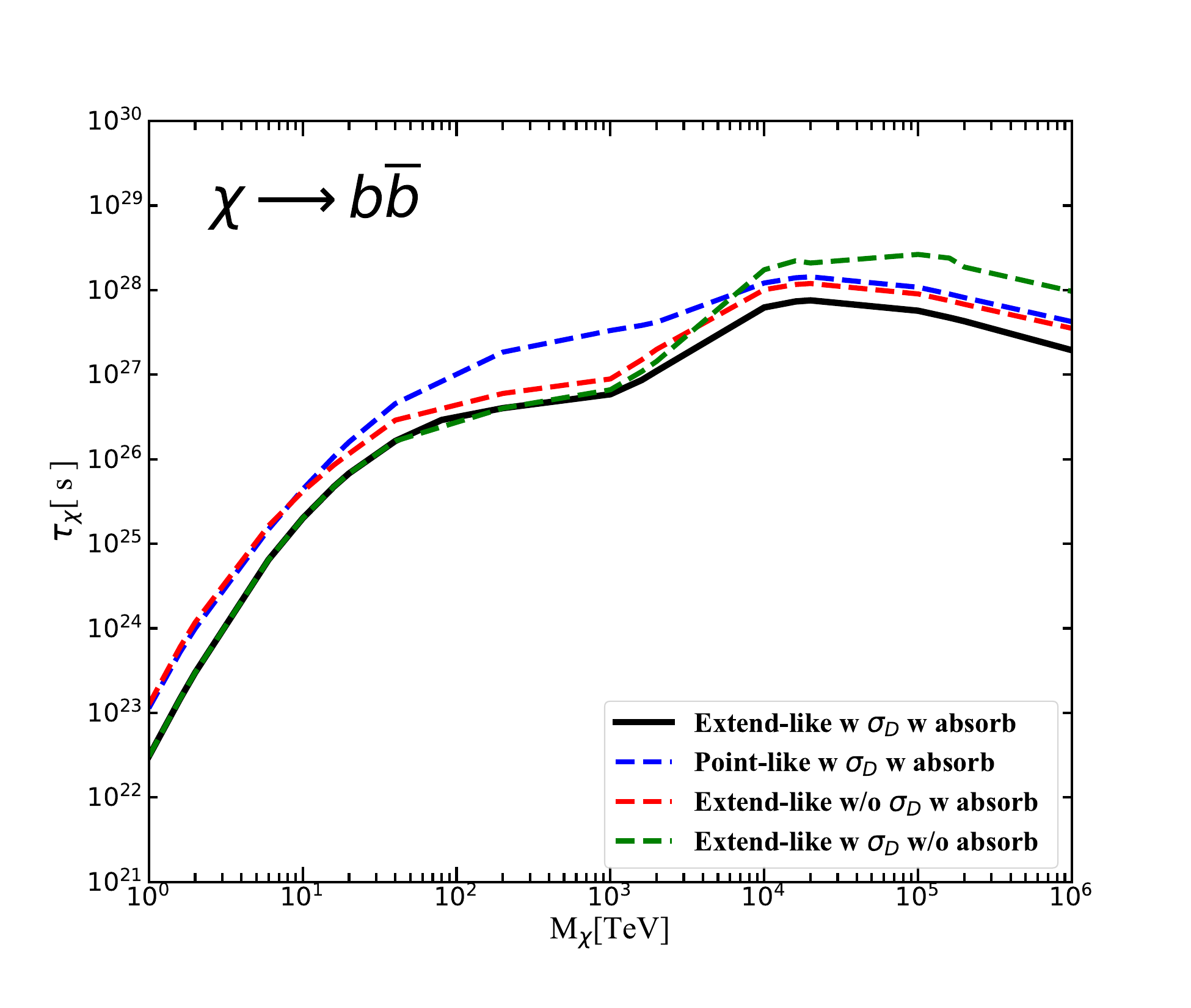}
                	\includegraphics[width=0.4\textwidth]{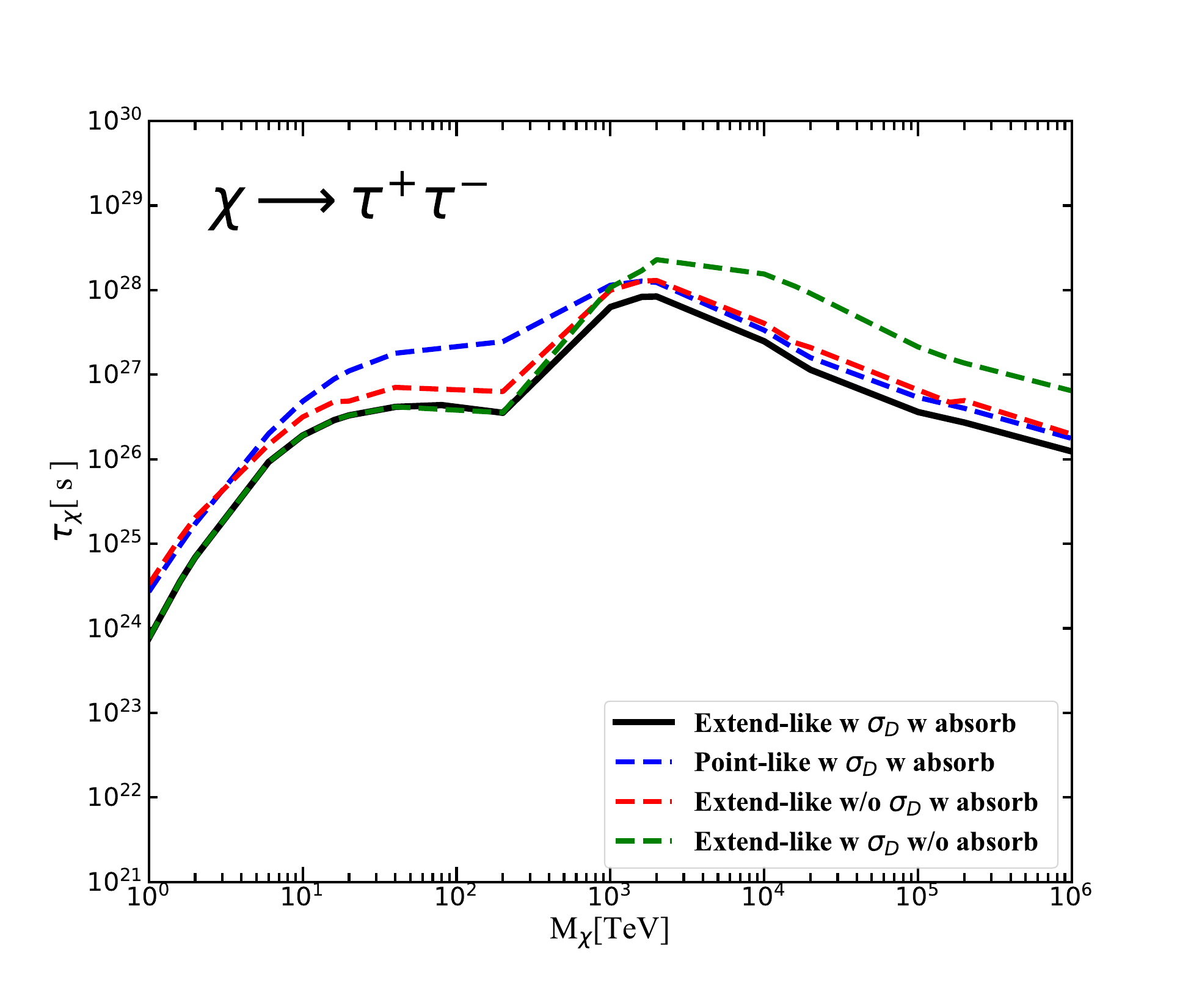}
                 \caption{
                Comparison of results for the impacts of high energy gamma-ray absorption by the Interstellar Radiation Field (ISRF), uncertainty of J- (D-) factor and morphology processing of dSphs on DM parameter constraints. The top panels represent the compared combined limits of DM annihilation cross-section of ${b\overline{b}}$ and ${\tau^{+}\tau^{-}}$ channels, while down panels represent the compared combined limits of DM decay lifetime. 
                The black line represents the limits of the extended source analysis with J- (D-) factor uncertainty and the effect of absorption, as the benchmark analysis. The dashed green line represents the limits of the extended source analysis considering only the J- (D-) factor uncertainty, and the dashed red line represents the results of the extended source analysis considering only the effect of absorption by ISRF. The dashed blue line represents the limits of the point-like source analysis with J- (D-) factor uncertainty and the effect of absorption by ISRF. }
                \label{Compare}
            \end{figure*}

In Fig~\ref{Compare}, we show comparison of results for the impacts of High Energy gamma-ray absorption by the Interstellar Radiation Field (ISRF), uncertainty of J- (D-) factor and morphology processing of dSphs on DM parameter constraints.

\begin{figure*}[htbp]
\includegraphics[height=0.31\textwidth,width=0.4\textwidth]{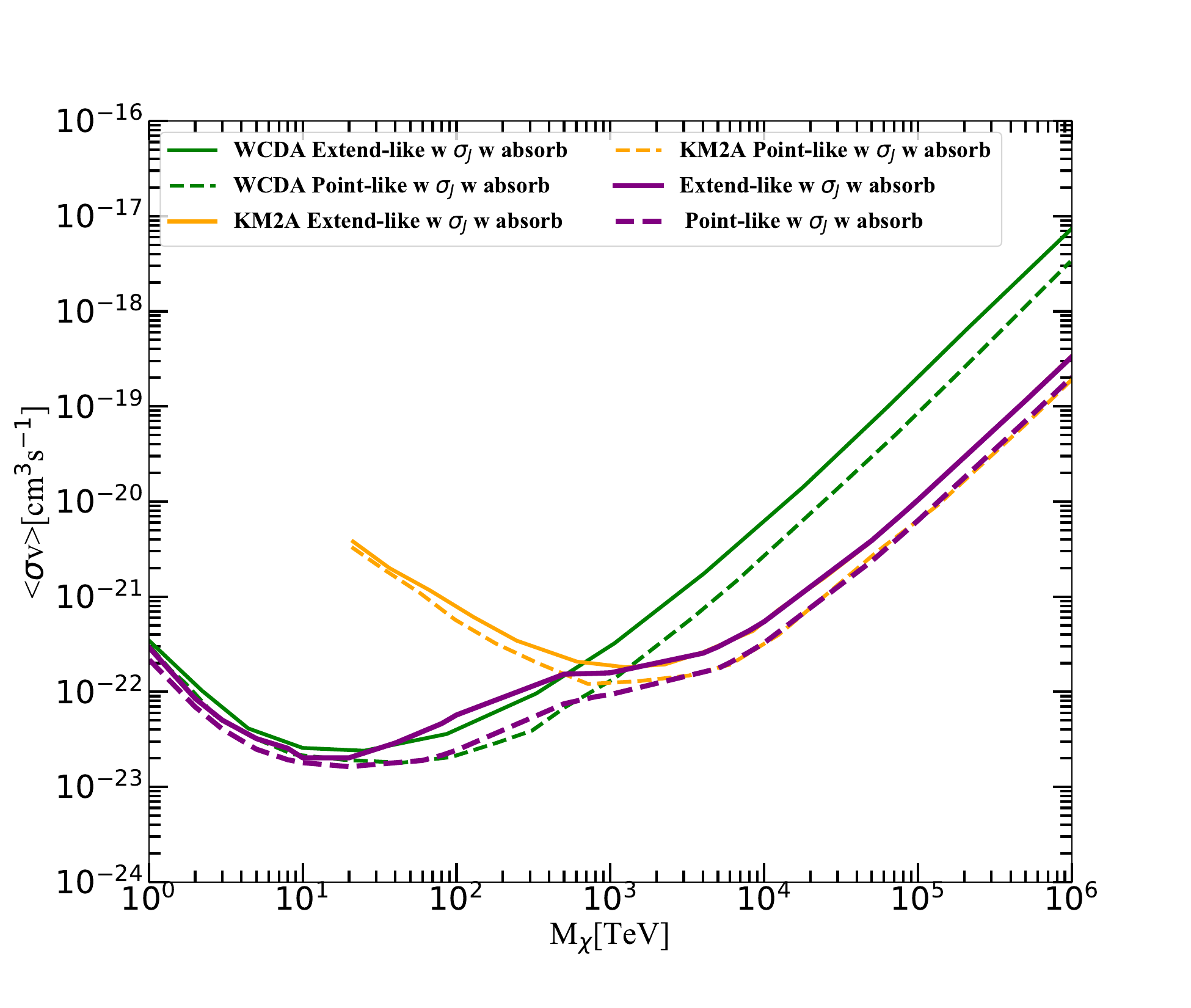}
\includegraphics[height=0.31\textwidth,width=0.4\textwidth]{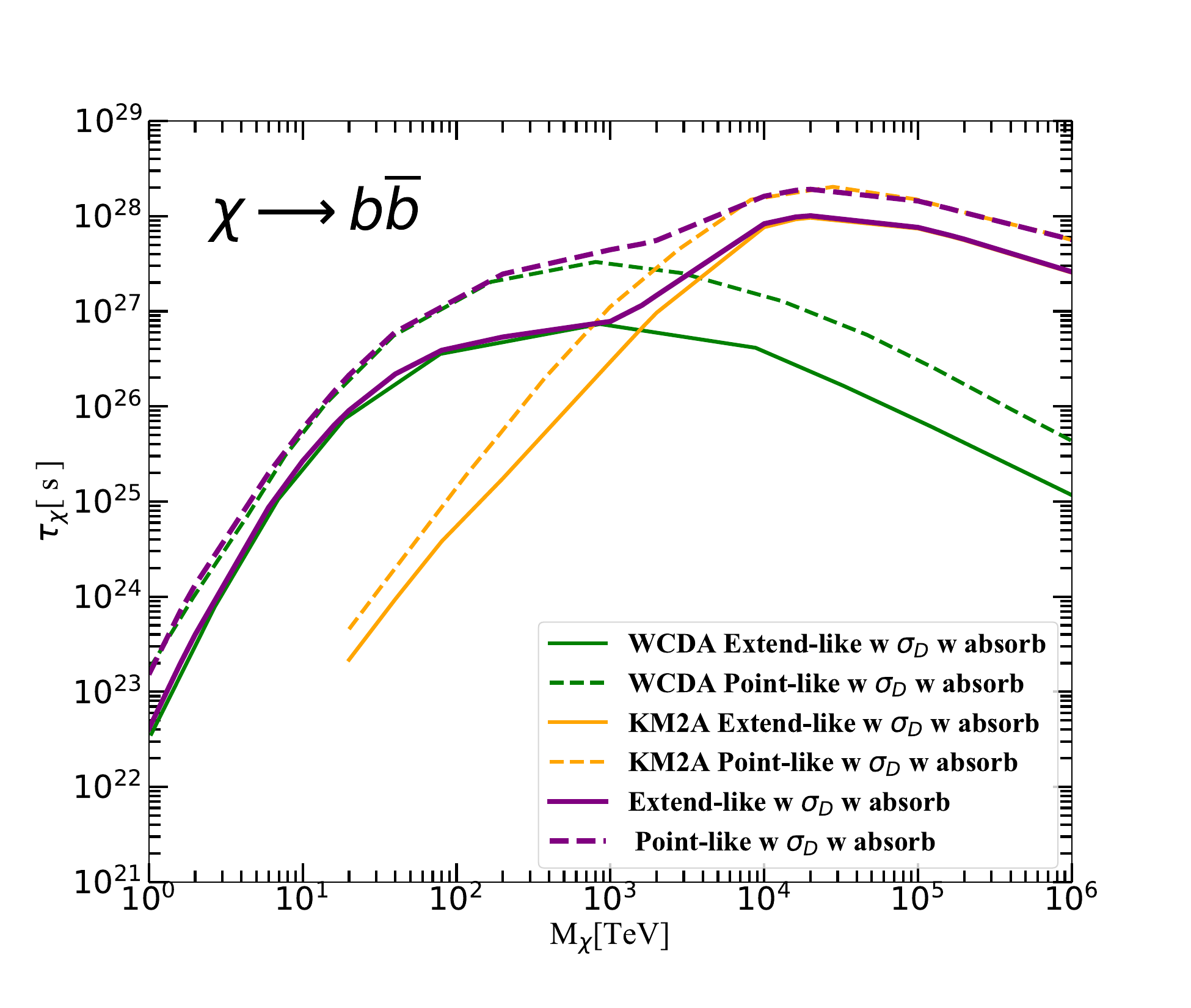}
\caption{Comparison of results for the impacts of treating dSphs as point sources versus extended sources, using WCDA data only, KM2A data only, and combined WCDA and KM2A data. The 95$\%$ C.L. upper limits on the DM annihilation cross-section for the $b\overline{b}$ channels are shown in the left panel. The 95$\%$ C.L. lower limits on the DM lifetime for the $b\overline{b}$ channels are presented in the right panel.}
\label{fig:WCDA_KM2A}
\end{figure*}

In Fig~\ref{fig:WCDA_KM2A}, we show comparison of results for the impacts of treating dSphs as point sources versus extended sources, using WCDA data only, KM2A data only, and combined WCDA and KM2A data. It is clear that, in the case of KM2A data, constraints on low-mass dark matter remain consistent between point source and extended source analyses, which is consistent with the large PSF at lower energy levels. However, constraints on high-mass dark matter are degraded significantly in the extended analysis due to the high-quality PSF at higher energy levels. Similar trends in constraints are observed in the results obtained from WCDA. As anticipated, the signal originating from decay processes should exhibit even greater spatial extension compared to that from annihilation. Thus it is evident that the good PSF of LHAASO could enable us to take the spatial analysis. 

\begin{figure*}[htbp]
	\includegraphics[height=0.4\textwidth   ,width=0.4\textwidth]{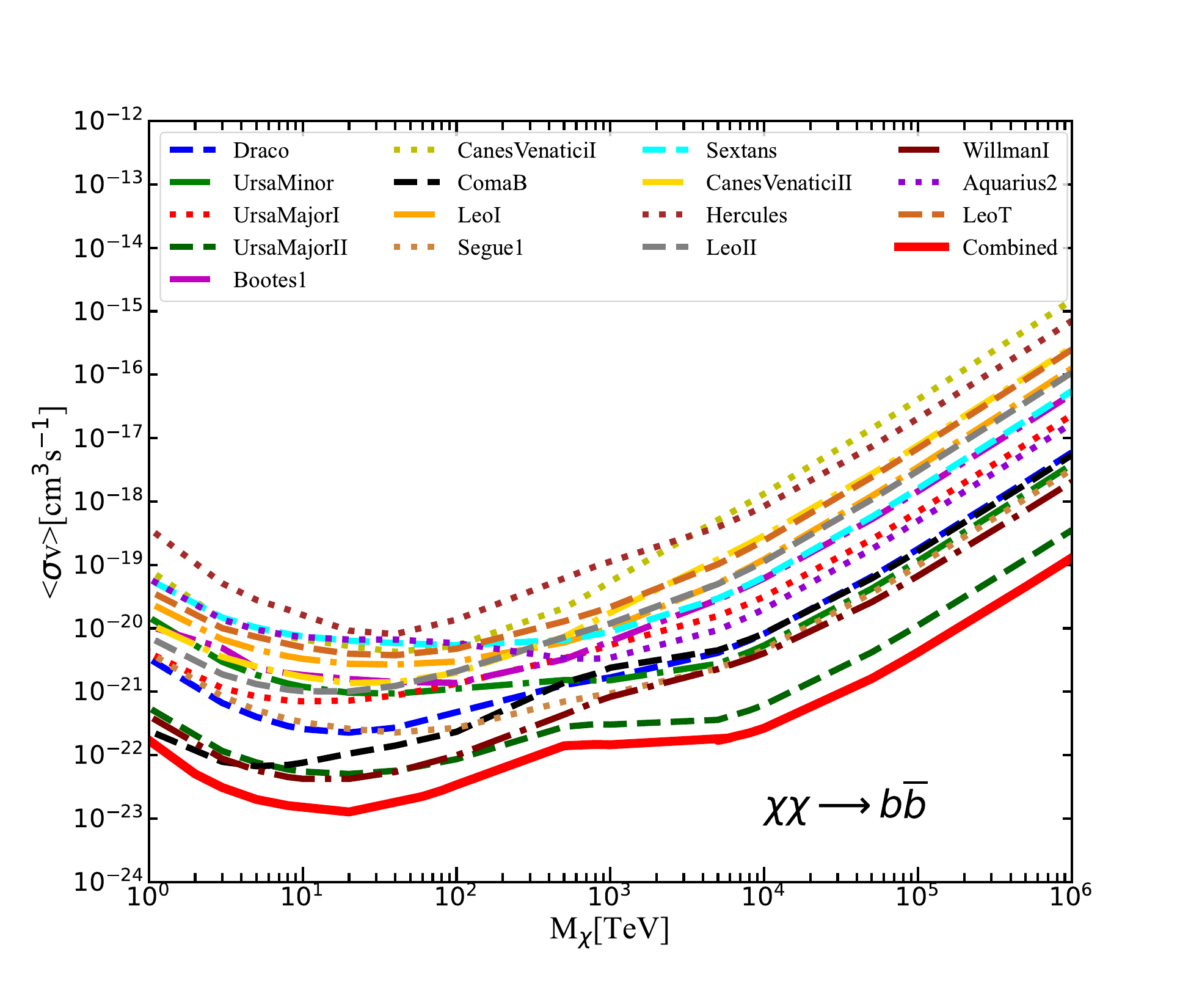}
	\includegraphics[height=0.4\textwidth,width=0.4\textwidth]{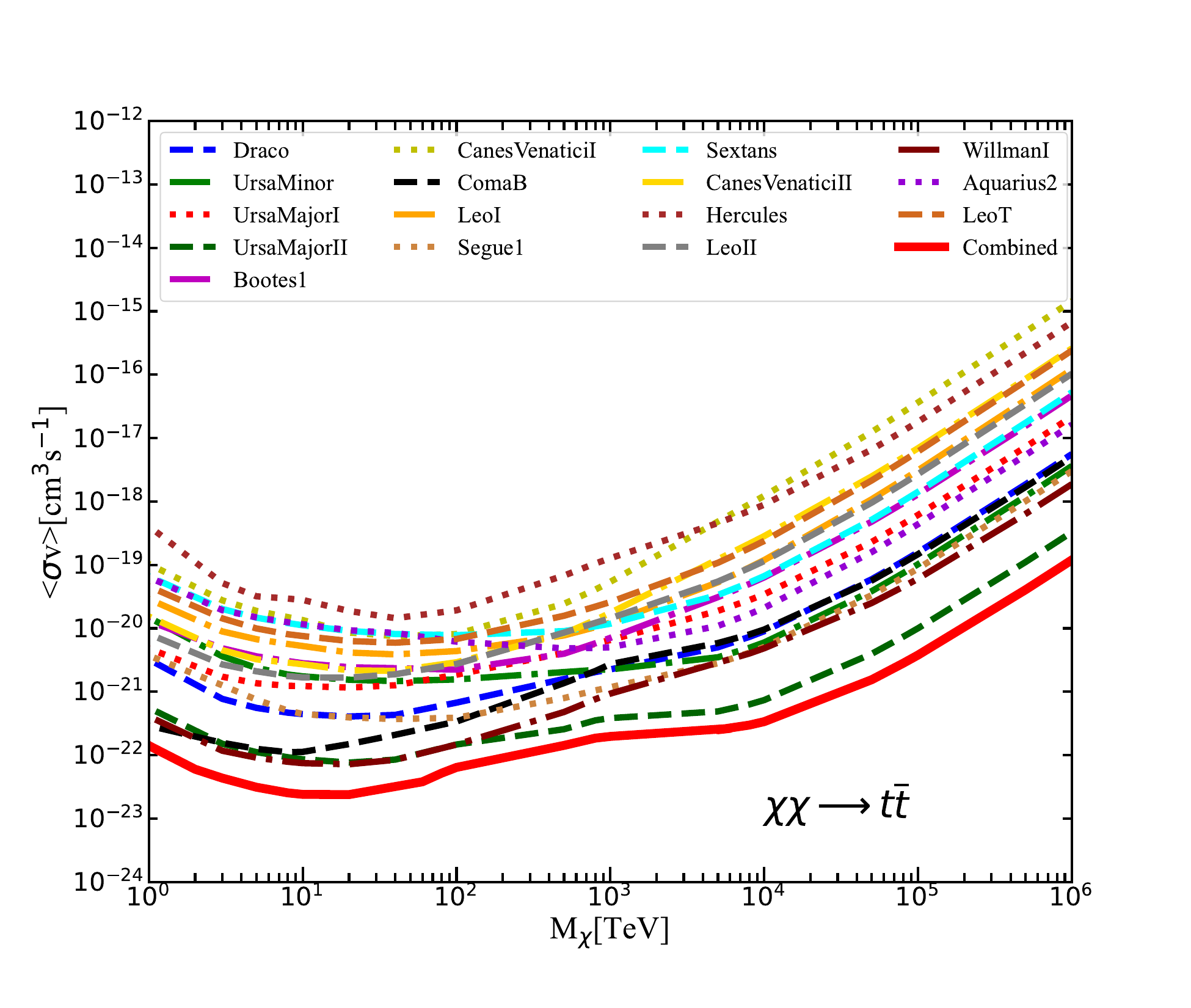}
	\includegraphics[height=0.4\textwidth,width=0.4\textwidth]{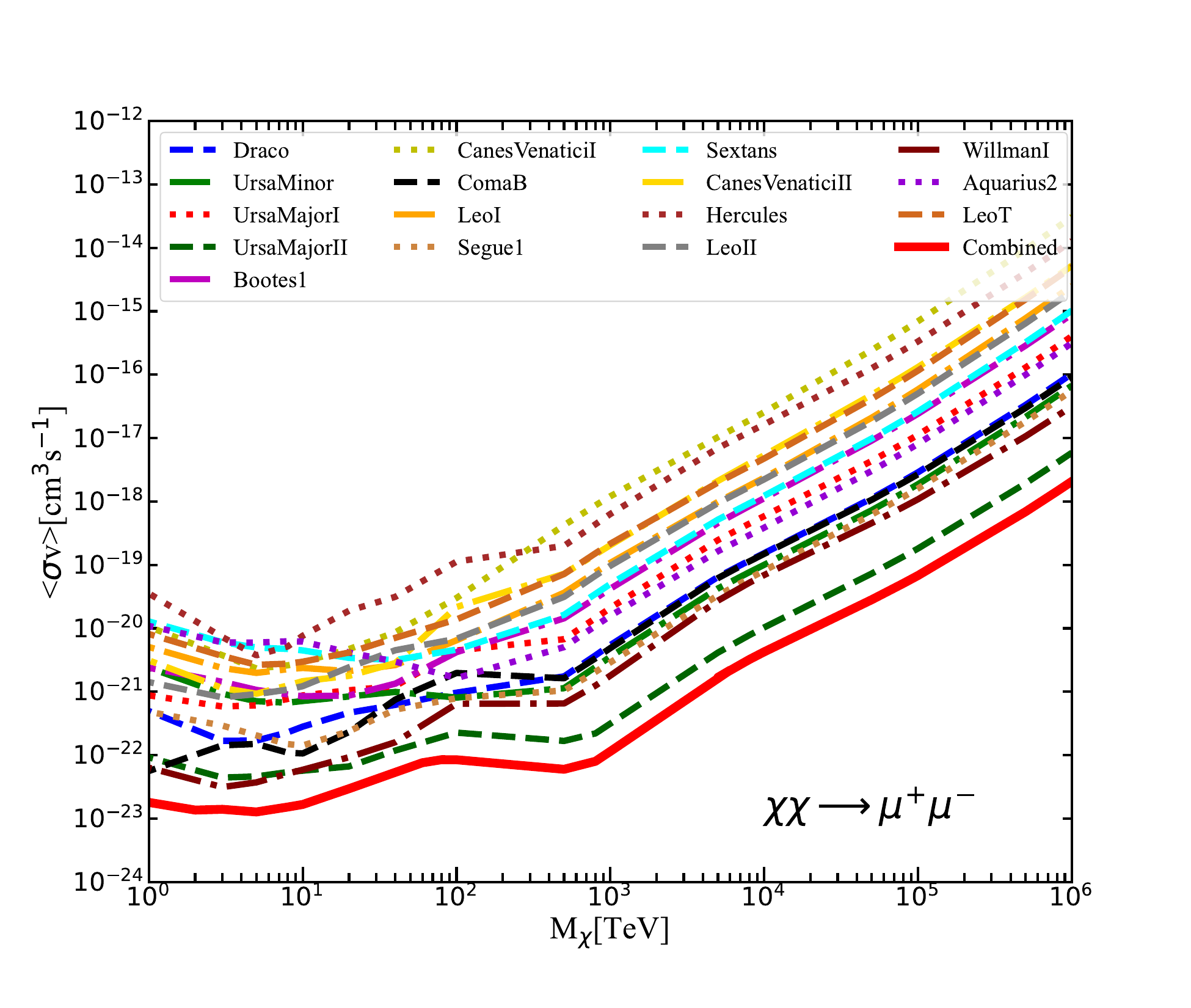}
	\includegraphics[height=0.4\textwidth,width=0.4\textwidth]{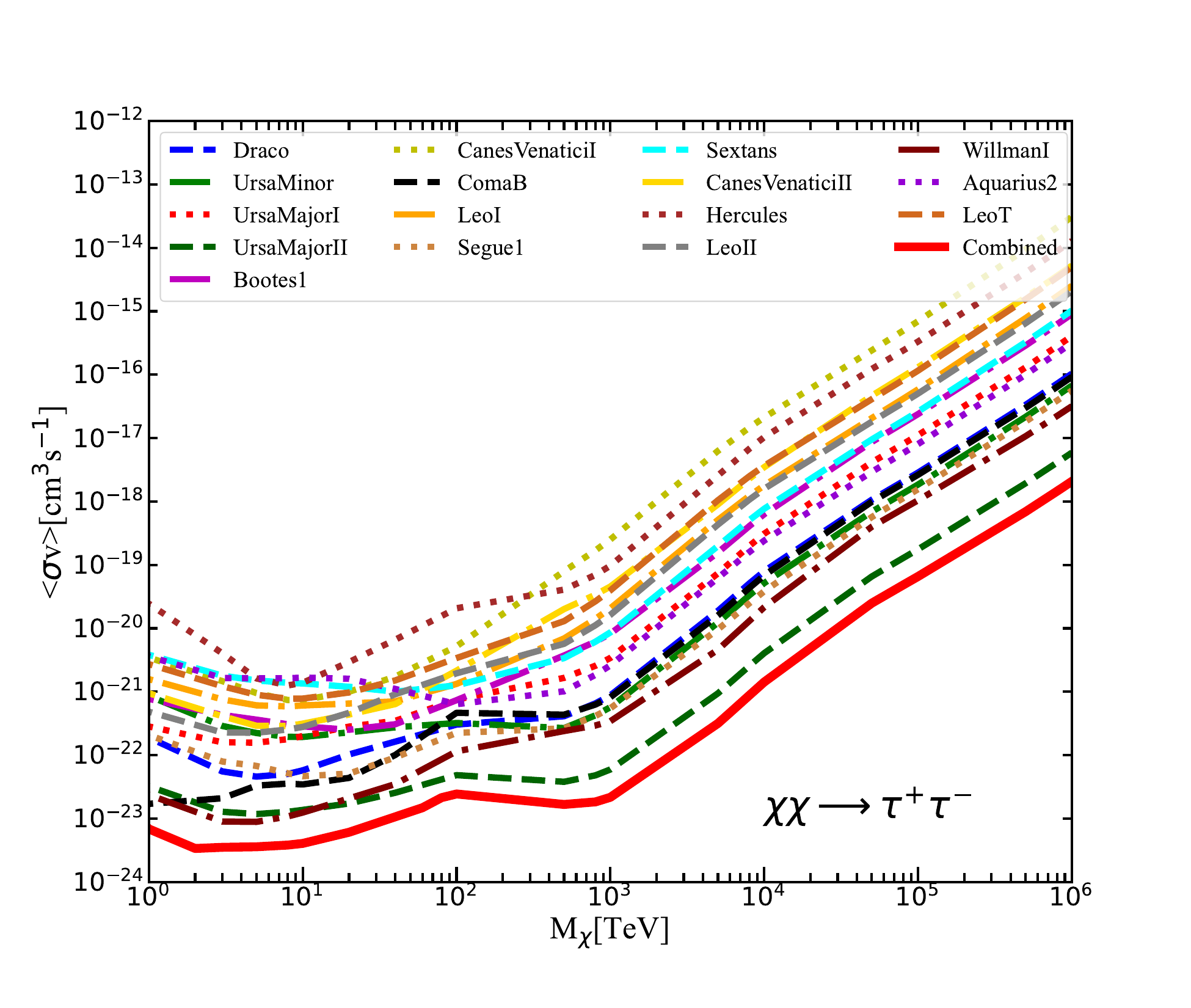}
	\includegraphics[height=0.4\textwidth,width=0.4\textwidth]{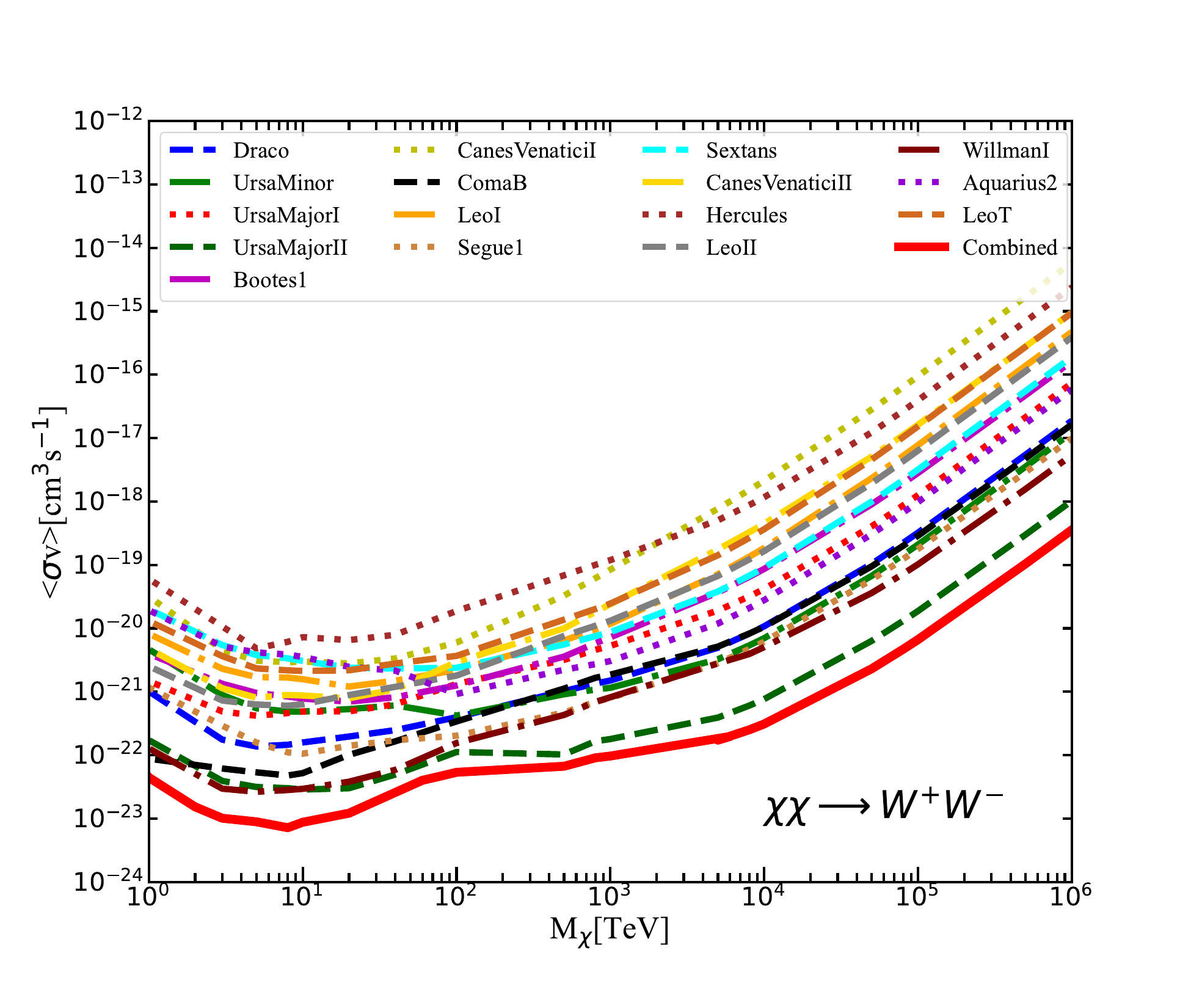}
	\caption{The results of LHAASO analysis show the 95$\%$ C.L. upper limits on the DM annihilation cross-section for five channels ($b\overline{b}$,$\tau^{+}\tau^{-}$, $t\overline{t}$,$\mu^{+}\mu^{-}$ and $W^{+}W^{-}$) using data from 16 dSphs around the Milky Way. The solid red line represents the combined limits obtained using all of the dSphs, while the dashed colored lines represent the individual upper limits obtained from each dSph.}
	\label{limits_anni_indicidual}
\end{figure*}

\begin{figure*}[!t]

	\includegraphics[height=0.4\textwidth,width=0.4\textwidth]{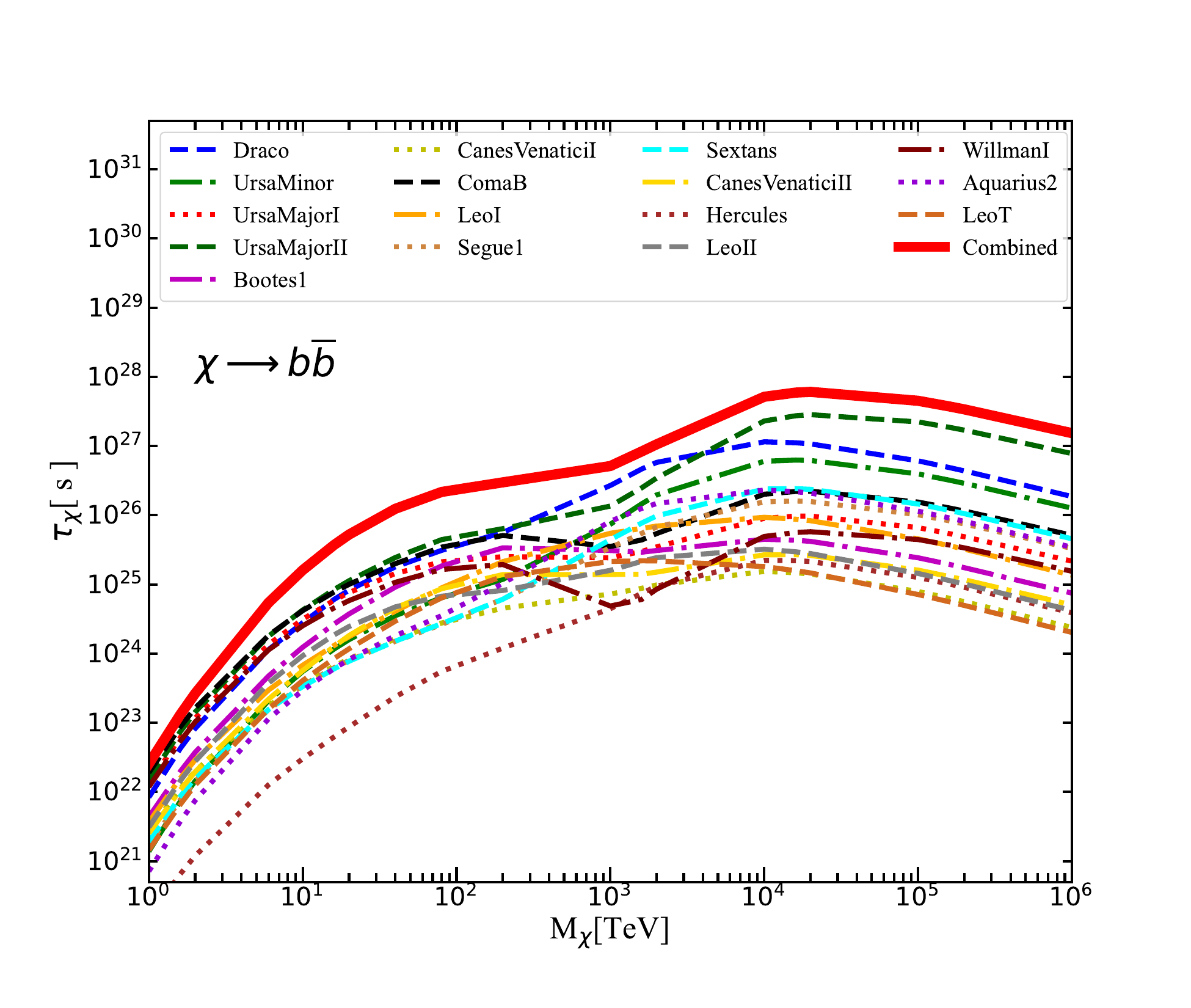}
	\includegraphics[height=0.4\textwidth,width=0.4\textwidth]{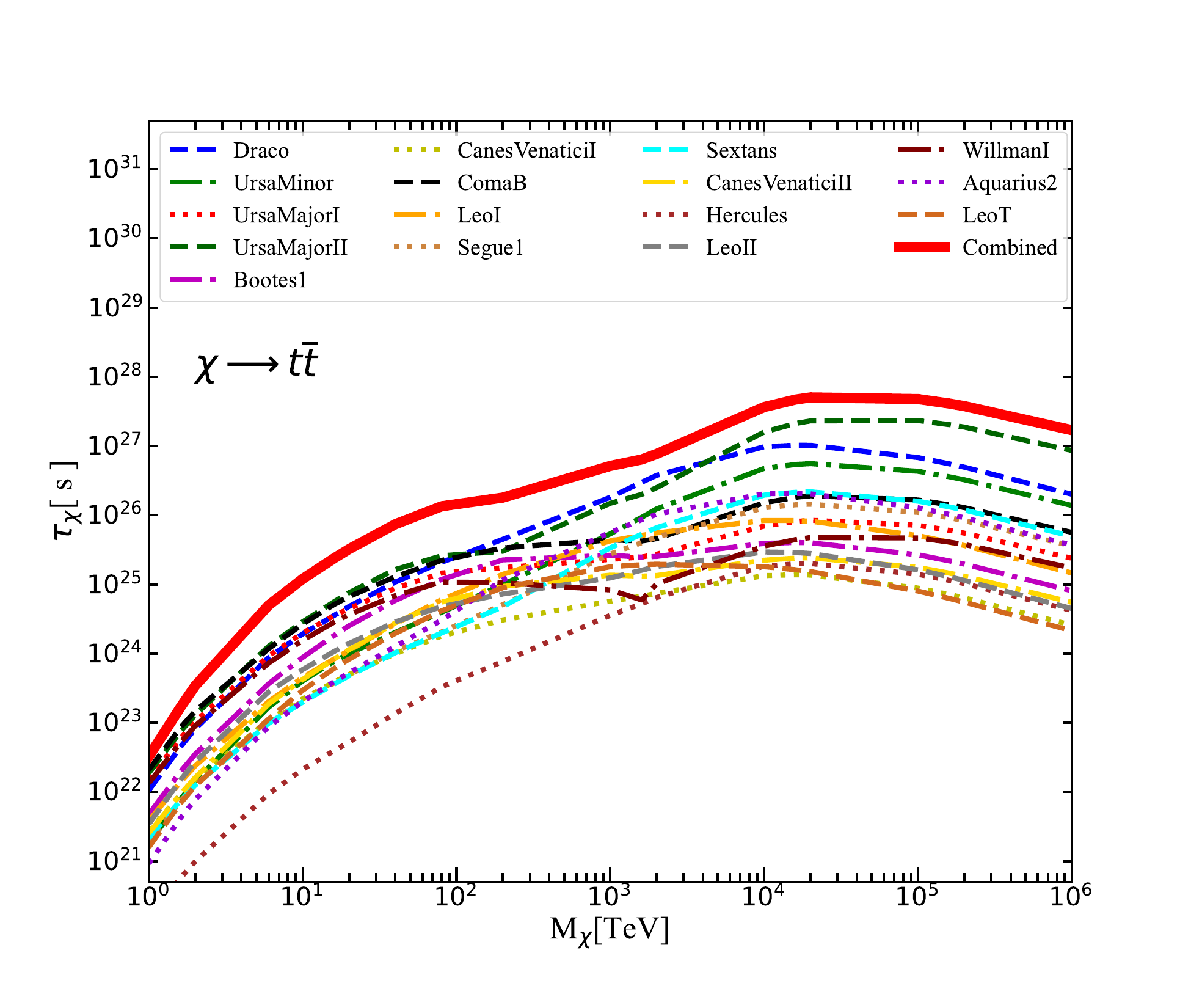}
	\includegraphics[height=0.4\textwidth,width=0.4\textwidth]{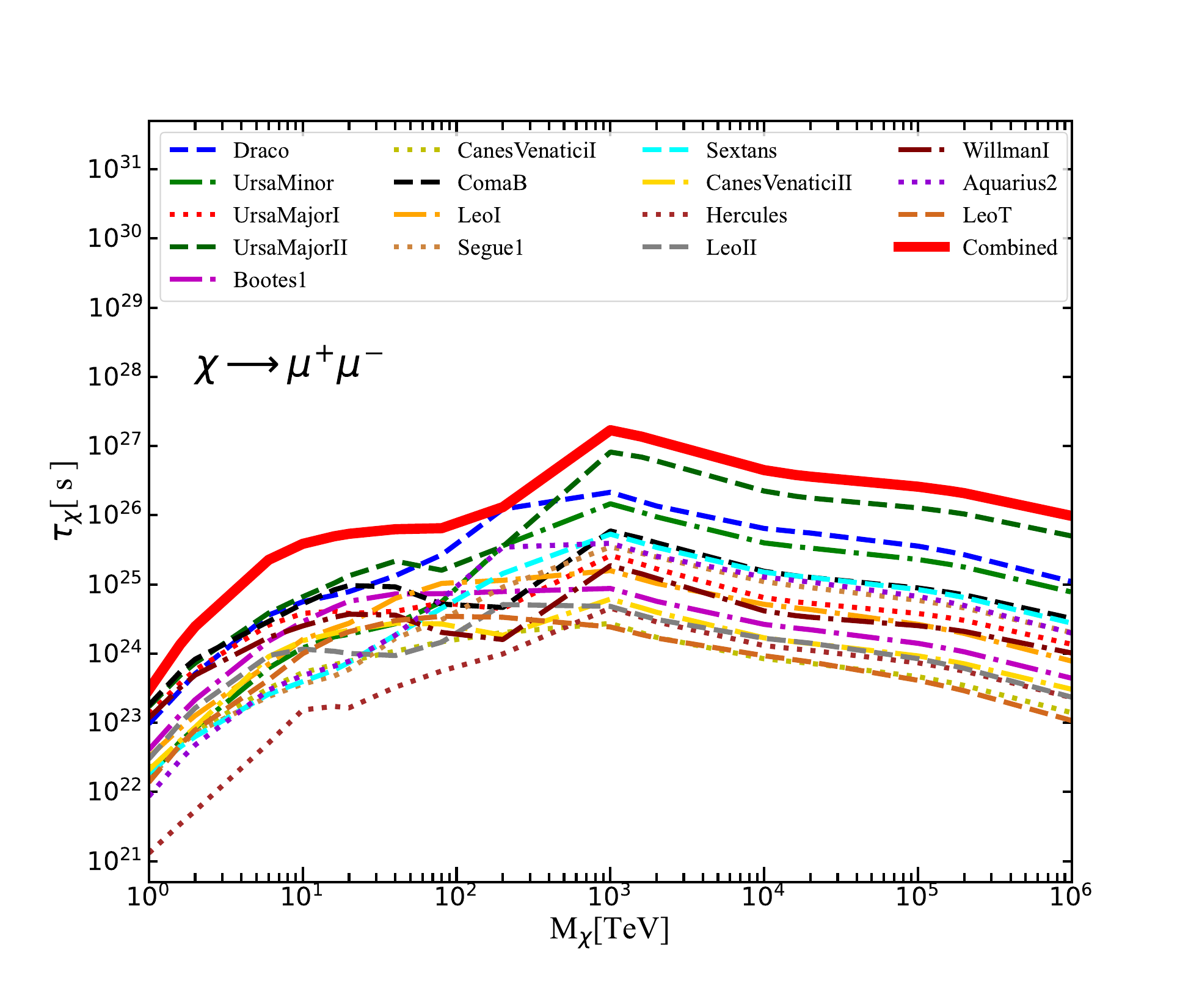}
	\includegraphics[height=0.4\textwidth,width=0.4\textwidth]{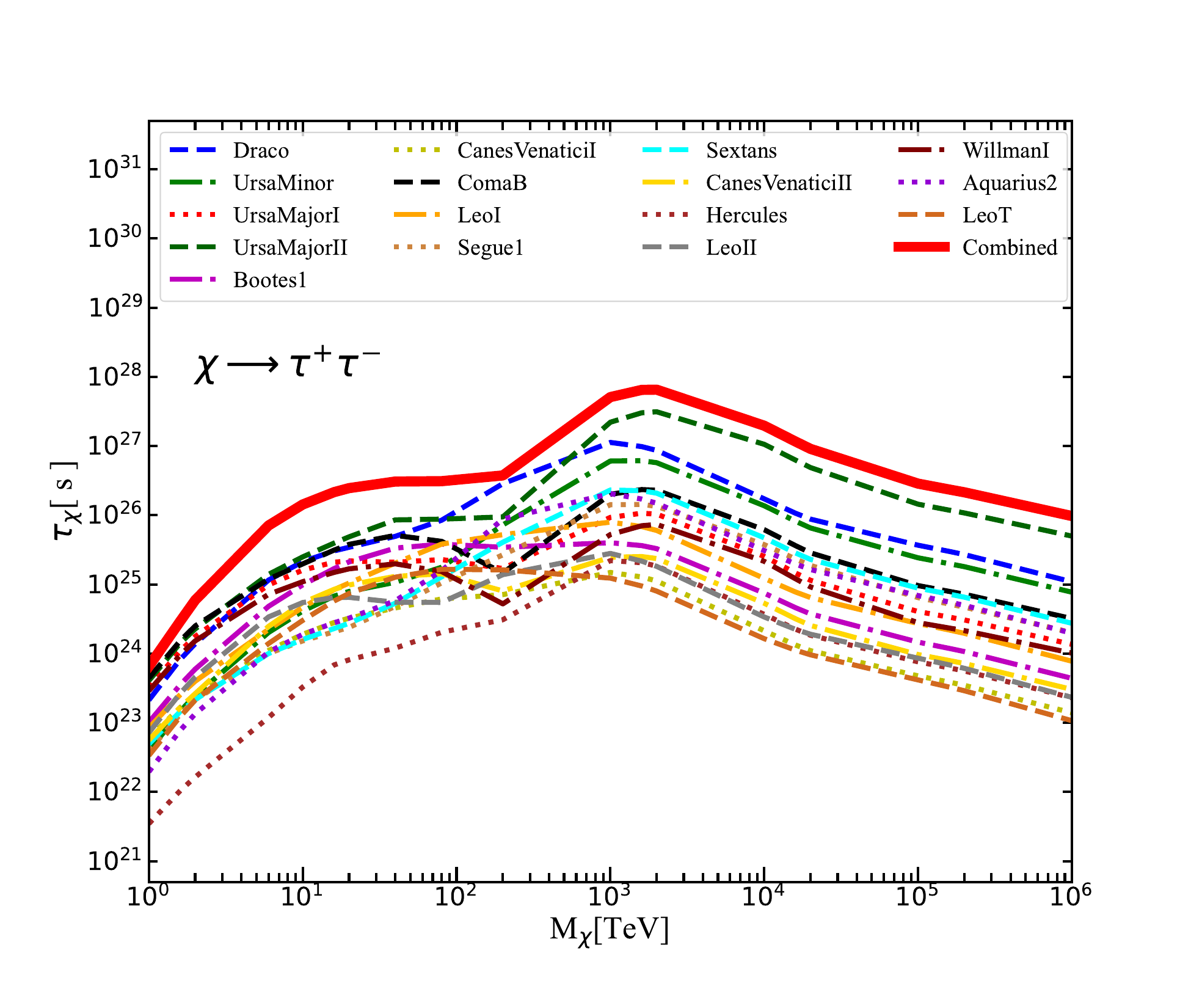}
	\includegraphics[height=0.4\textwidth,width=0.4\textwidth]{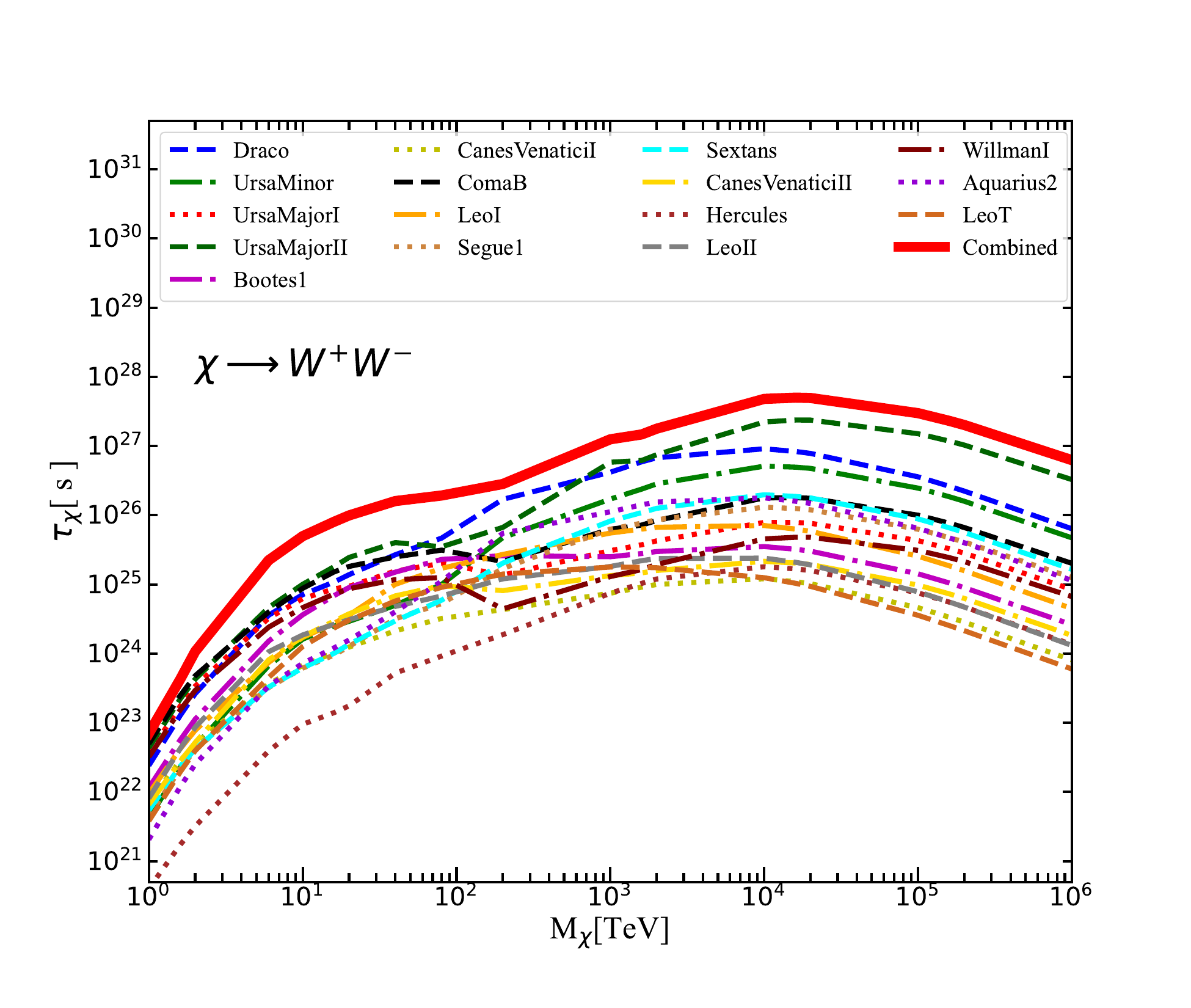}
	
	\caption{The panels display the 95$\%$ C.L. lower limits on the DM decay lifetime assuming DM with mass range from 1 TeV to 1 EeV with 100$\%$ branching ratio decay into specific standard model particles($b\overline{b}$,$\tau^{+}\tau^{-}$, $t\overline{t}$,$\mu^{+}\mu^{-}$ and $W^{+}W^{-}$) for all 16 dSphs. The solid red line represents the combined limit obtained from the analysis of all dSphs, while the dashed colored lines correspond to the lower limits obtained from the individual analysis of each dSph. }
	\label{limits_decay_indicidual}
\end{figure*}

In Fig~\ref{limits_anni_indicidual} and Fig~\ref{limits_decay_indicidual}, we show 95$\%$ C.L. upper limits on the DM annihilation cross-section and 95$\%$ C.L. lower limits on the DM decay lifetime from individual analysis of each dSph and combined analysis of all dSphs.

\begin{figure*}[htbp]
                 	%\begin{left}

                 \includegraphics[height=0.28\textwidth,width=0.32\textwidth]{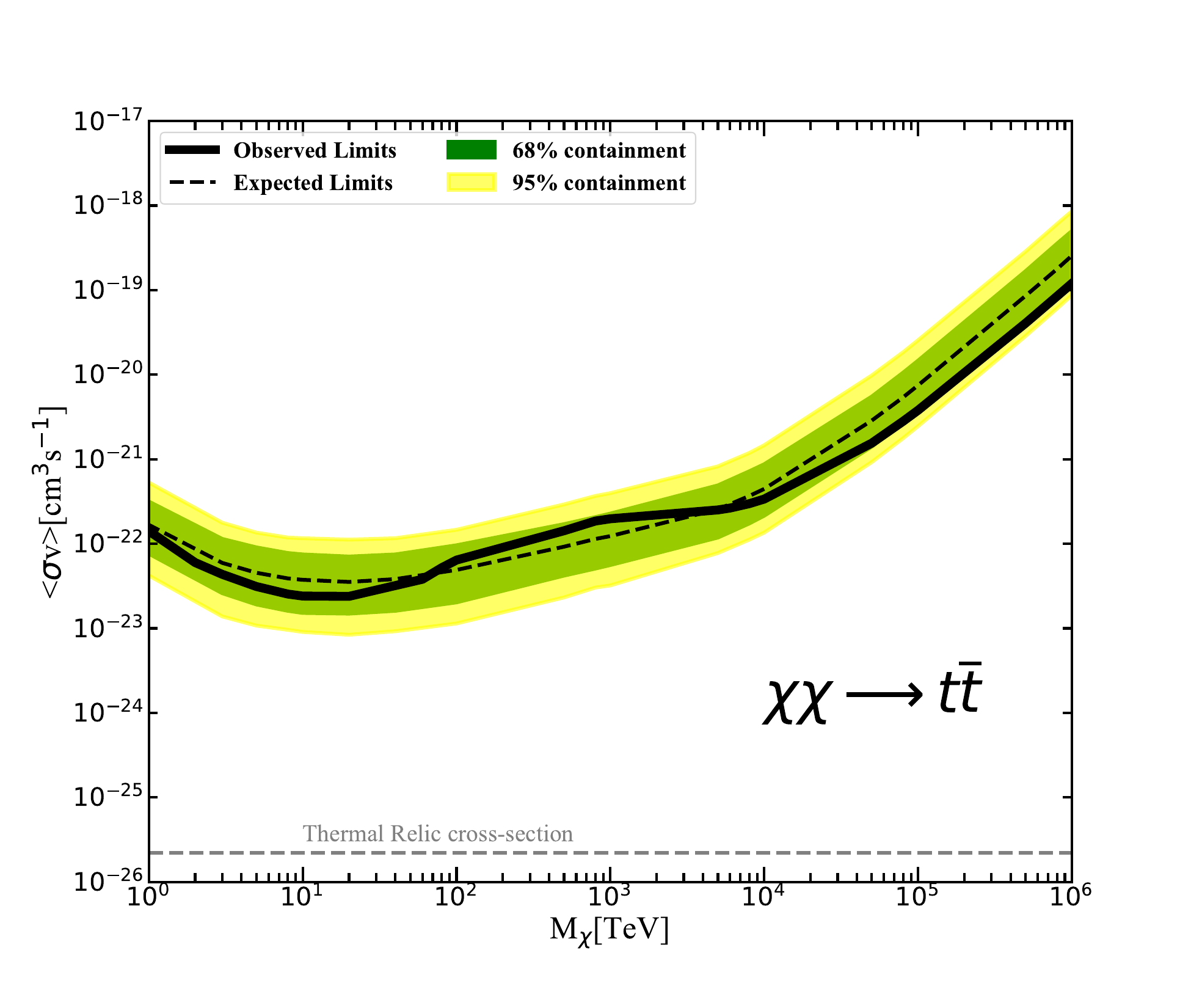}
                 	\includegraphics[height=0.28\textwidth,width=0.32\textwidth]{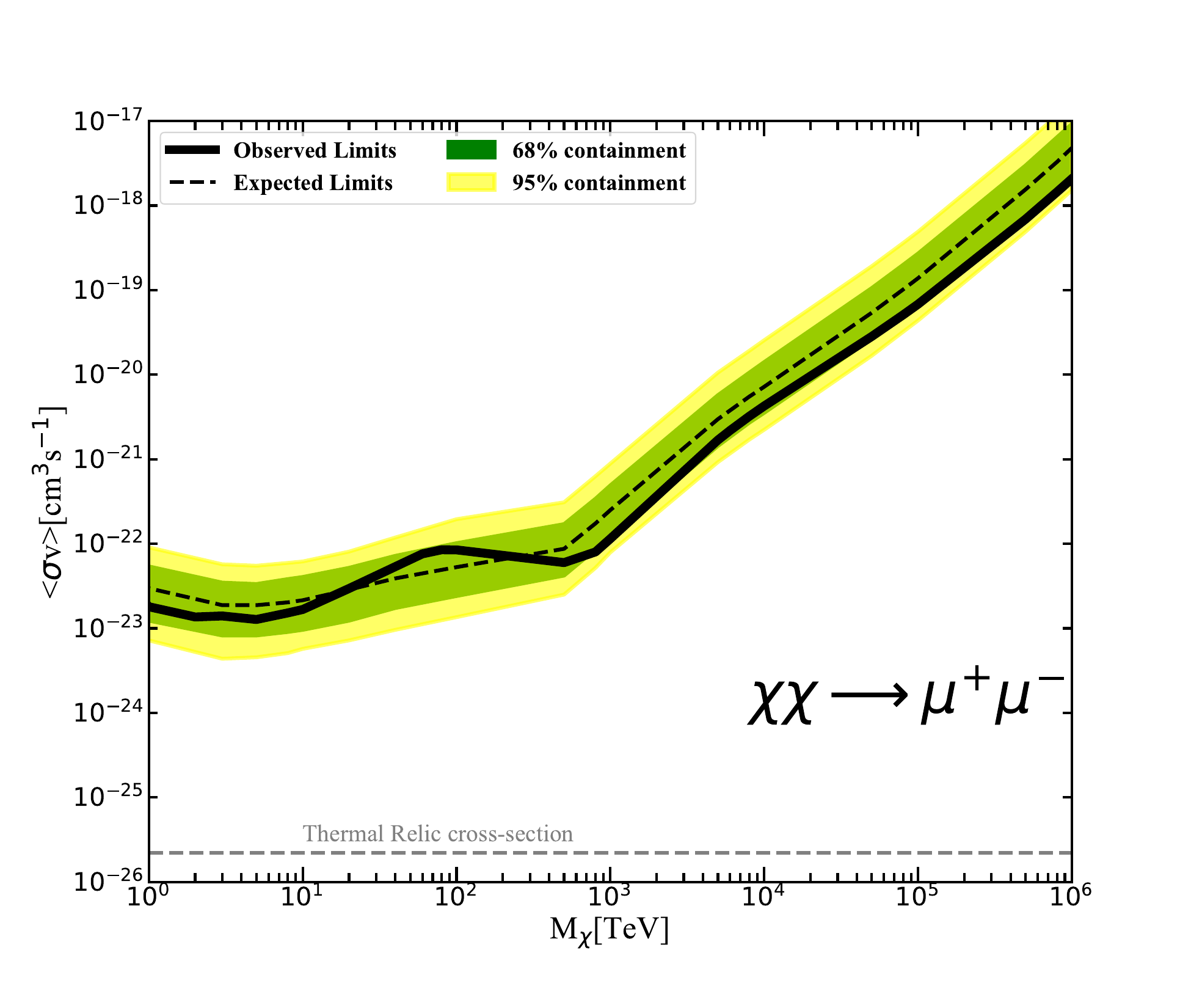}
                 	\includegraphics[height=0.28\textwidth,width=0.32\textwidth]{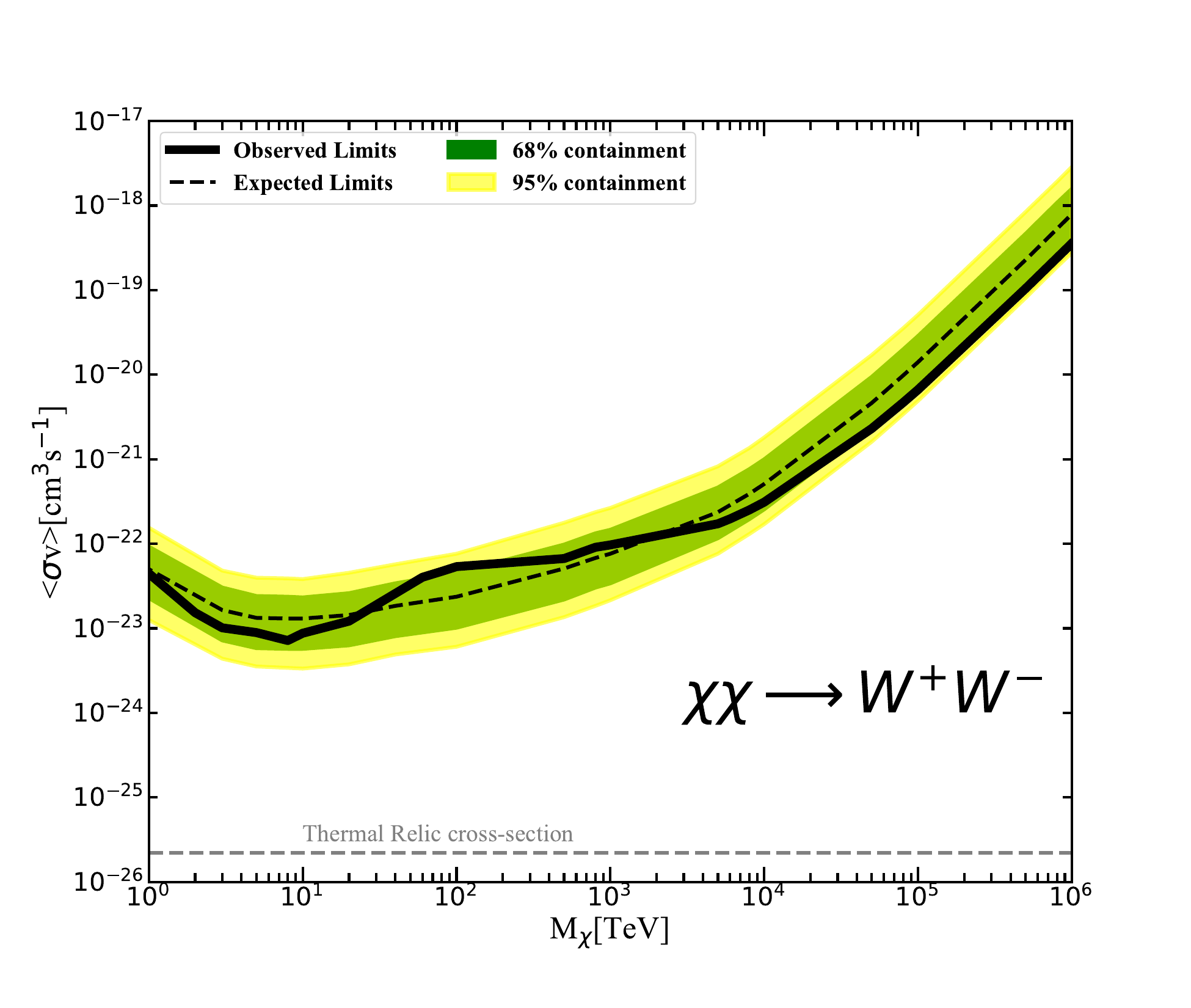}
                 \includegraphics[height=0.28\textwidth,width=0.32\textwidth]{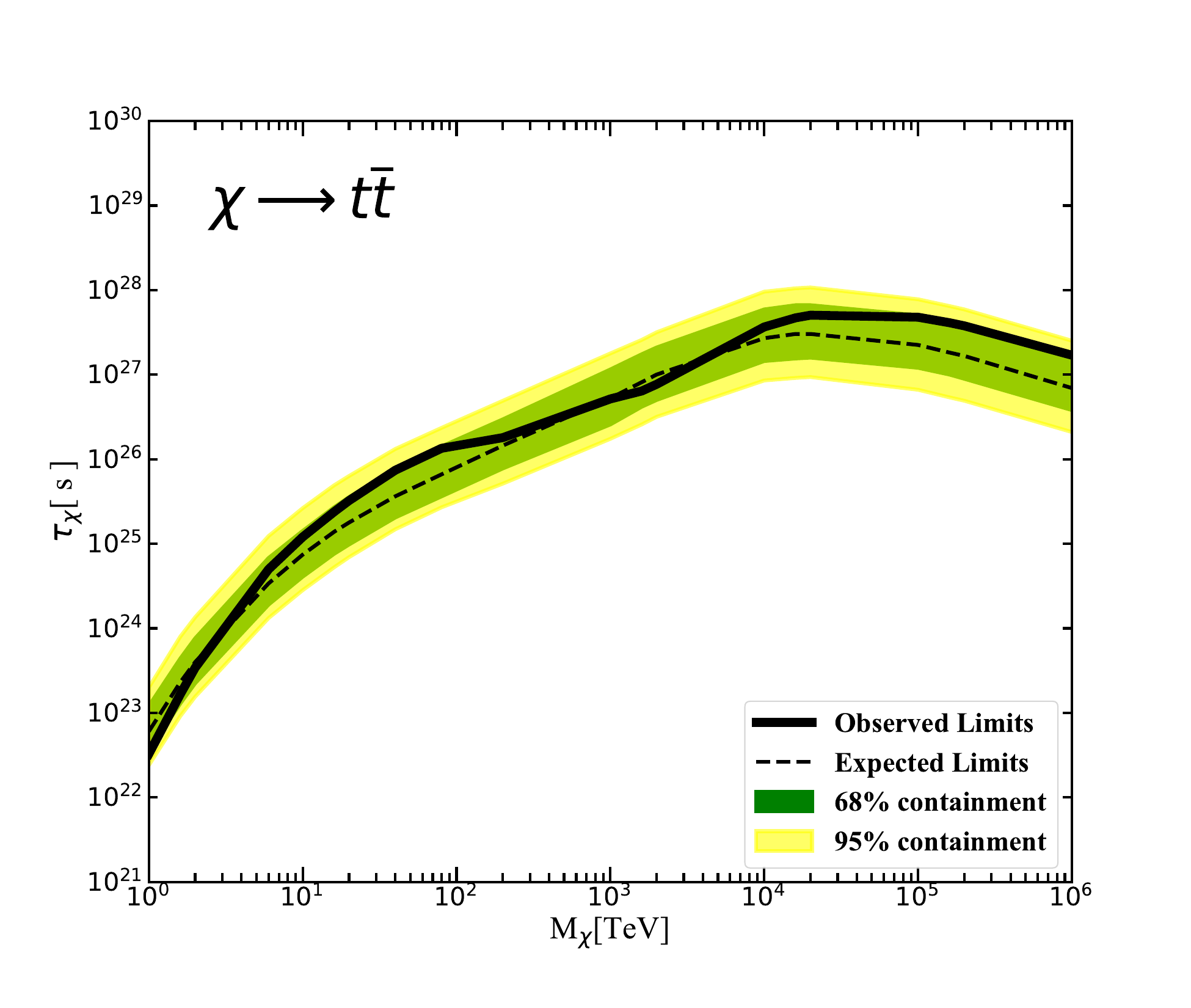}
                 \includegraphics[height=0.28\textwidth,width=0.32\textwidth]{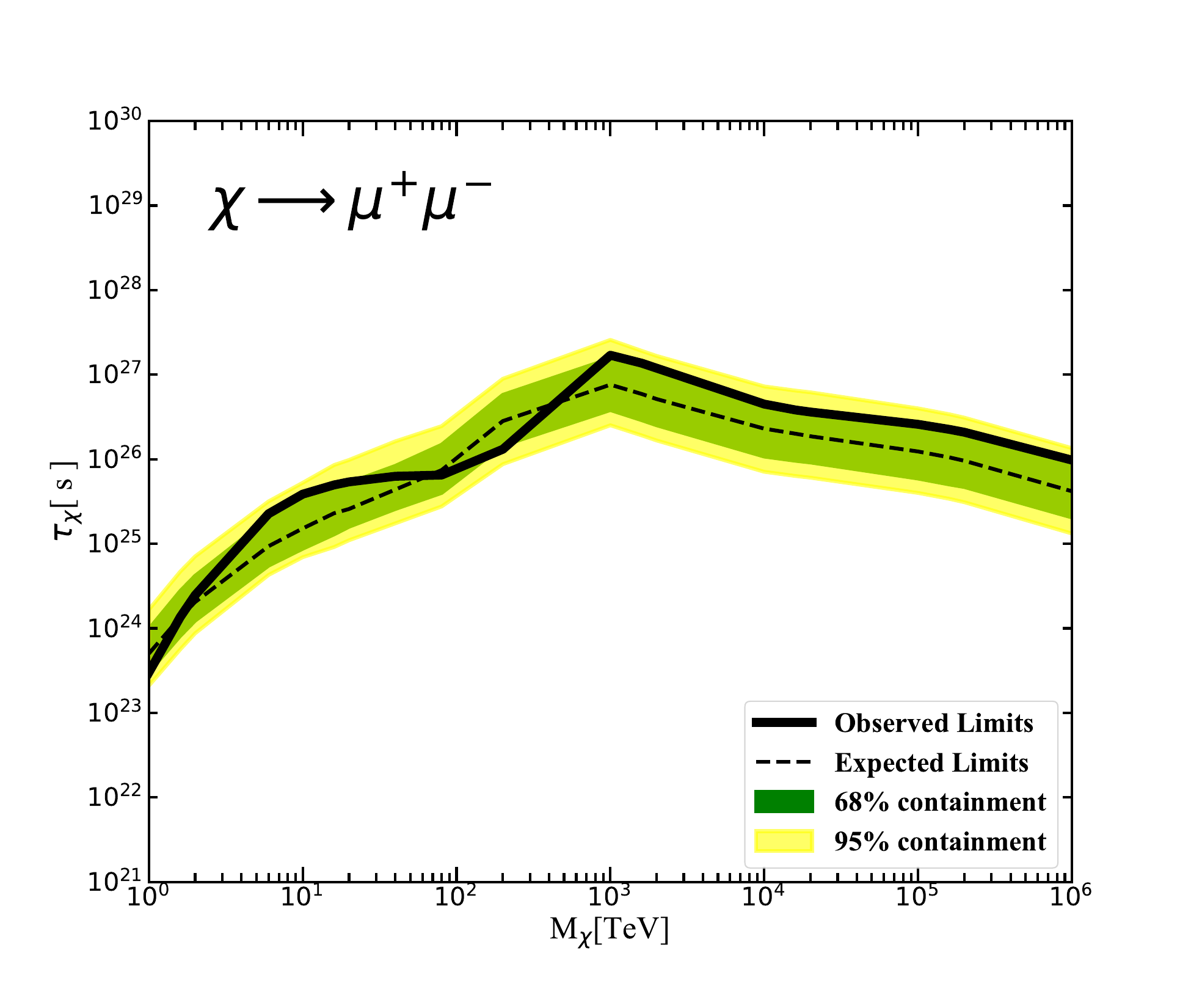}
                 \includegraphics[height=0.28\textwidth,width=0.32\textwidth]{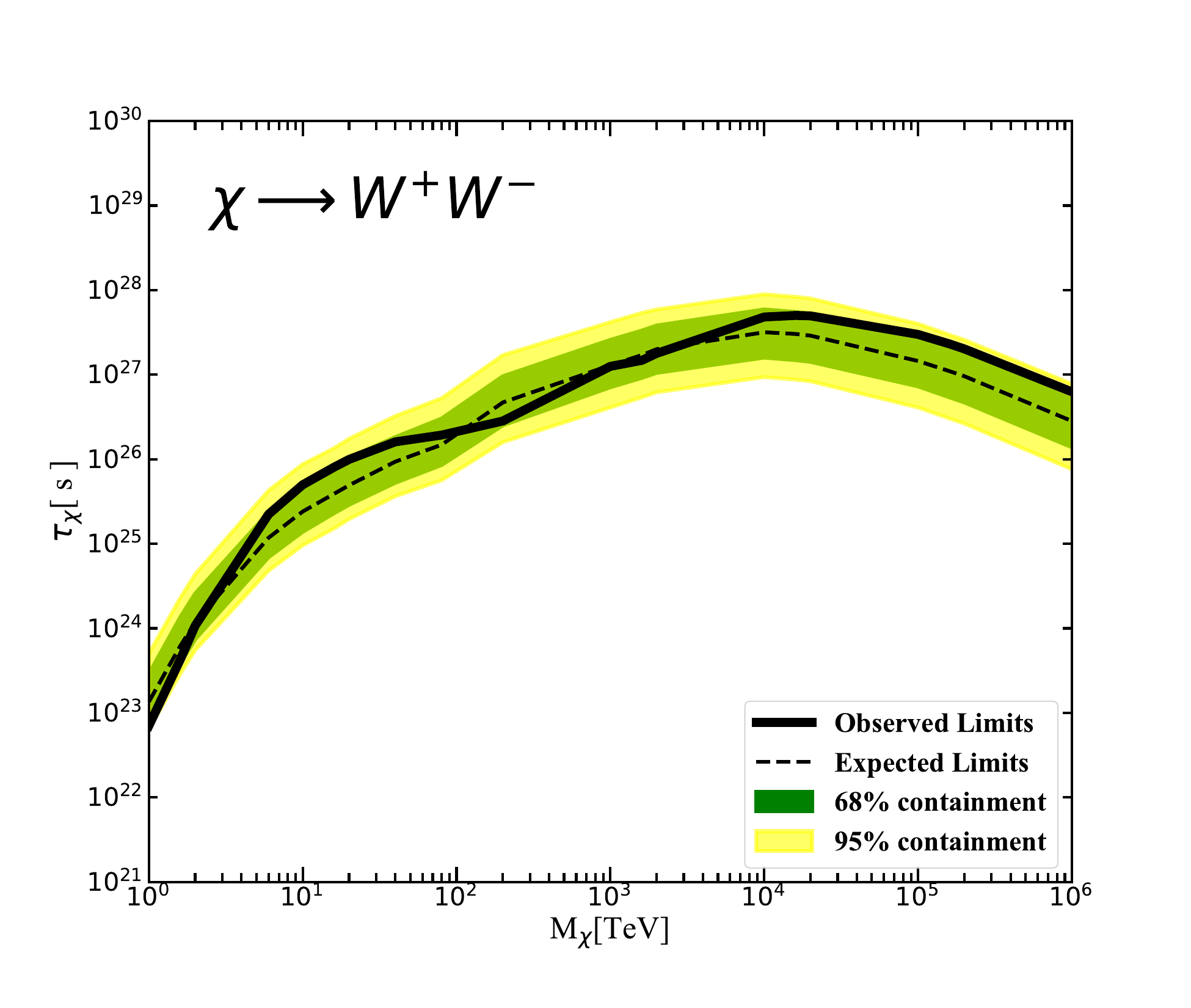}
                 \\	
            	\caption{The comparison of observed limits and expected limits for the $t\overline{t},\mu^{+}\mu^{-}$ and $W^{+}W^{-}$ processes. The top three figures depict the observed and expected 95$\%$ C.L. upper limits on the DM annihilation cross-section. The bottom three figures represent the observed 95$\%$ C.L. lower limits on the DM decay lifetime. The solid black line represents the LHAASO observed combined limit, and the dashed black line, green band, and yellow band represent the expected combined limits, their 1$\sigma$ and 2$\sigma$ uncertainty based on the mock observations.}
            	\label{limits_expected}
            \end{figure*}

In Fig~\ref{limits_expected}, we show the comparison of observed limits and expected limits from pure background simulation.

%\FloatBarrier         

%TC:endignore
\end{document}